\documentclass[journal]{IEEEtran}

\usepackage{xcolor,soul,framed}

\ifCLASSINFOpdf
   \usepackage[pdftex]{graphicx}
   \graphicspath{{../pdf/}{../jpeg/}}
   \DeclareGraphicsExtensions{.pdf,.jpeg,.png}
\else
   \usepackage[dvips]{graphicx}
   \graphicspath{{../eps/}}
   \DeclareGraphicsExtensions{.eps}
\fi
\usepackage{amsmath}
\usepackage{
algorithm,
algpseudocode
}
\usepackage{booktabs}

\ifCLASSOPTIONcompsoc
 \usepackage[caption=false,font=normalsize,labelfont=sf,textfont=sf]{subfig}
\else
 \usepackage[caption=false,font=footnotesize]{subfig}
\fi
\begin{document}
\title{A Two-layer Approach for Estimating Behind-the-Meter PV Generation Using Smart Meter Data}

% \author{Michael~Shell,~\IEEEmembership{Member,~IEEE,}
%         John~Doe,~\IEEEmembership{Fellow,~OSA,}
%         and~Jane~Doe,~\IEEEmembership{Life~Fellow,~IEEE}% <-this % stops a space
% \thanks{M. Shell was with the Department
% of Electrical and Computer Engineering, Georgia Institute of Technology, Atlanta,
% GA, 30332 USA e-mail: (see http://www.michaelshell.org/contact.html).}% <-this % stops a space
% \thanks{J. Doe and J. Doe are with Anonymous University.}% <-this % stops a space
% \thanks{Manuscript received April 19, 2005; revised August 26, 2015.}}

\author{Fankun Bu,~\IEEEmembership{Graduate Student Member,~IEEE,}
        ~Rui Cheng,~\IEEEmembership{Graduate Student Member,~IEEE,}
        and Zhaoyu Wang,~\IEEEmembership{Senior Member,~IEEE}
\thanks{This work was supported in part by the National Science Foundation under EPCN 2042314 and  in part by the Grid Modernization Initiative of the U.S. Department of Energy (DOE) under GMLC project 2.1.1 – FASTDERMS. (\textit{Corresponding author: Zhaoyu Wang})}
\thanks{F. Bu, R. Cheng and Z. Wang are with the Department of
Electrical and Computer Engineering, Iowa State University, Ames, IA 50011, USA (e-mail: fbu@iastate.edu; wzy@iastate.edu).}
}


\markboth{}%
{Shell \MakeLowercase{\textit{et al.}}: Bare Demo of IEEEtran.cls for IEEE Journals}



\maketitle

\begin{abstract}
As the cost of the residential solar system decreases, rooftop photovoltaic (PV) has been widely integrated into distribution systems. Most rooftop PV systems are installed behind-the-meter (BTM), i.e., only the net demand is metered, while the native demand and PV generation are not separately recorded. Under this condition, the PV generation and native demand are invisible to utilities, which brings challenges for optimal distribution system operation and expansion. In this paper, we have come up with a novel two-layer approach to disaggregate the unknown PV generation and native demand from the known hourly net demand data recorded by smart meters: 1) At the aggregate level, the proposed approach separates the total PV generation and native demand time series from the total net demand time series for customers with PVs. 2) At the customer level, the separated aggregate-level PV generation is allocated to individual PVs. These two layers leverage the spatial correlations of native demand and PV generation, respectively. One primary advantage of our proposed approach is that it is more independent and practical compared to previous works because it does not require PV array parameters, meteorological data and previously recorded solar power exemplars. This paper has verified our proposed approach using real native demand and PV generation data.
\end{abstract}


% \fi
\IEEEpeerreviewmaketitle

\section{Introduction} \label{sec:intro}

\IEEEPARstart{I}{N} the last decade, residential rooftop photovoltaic (PV) has been proliferating in distribution systems.  In most cases, utilities only install a bi-directional smart meter to record the net demand of customers with PVs. This type of installation is referred to as behind-the-meter (BTM), in which case the net demand equals native demand minus PV generation. Therefore, the PV generation produced by solar array and the native demand consumed by appliances are unknown to utilities. Only metering the net demand can reduce the financial cost for utilities; however, as the penetration level of PV increases, the unobservability of notable PV generation and native demand brings significant challenges to distribution systems. We focus on three specific applications to elaborate the necessity of estimating the unknown BTM PV generation and native demand: \textit{First}, the unavailability of native load and PV generation might cause unacceptable forecasting errors because some forecasters require reconstituting the generation and native demand time series \cite{forecasting_overview, kangping_recom_1}. In contrast, knowing BTM PV generation and native load can help utilities forecast generation and load separately, thus provide utilities useful information regarding load/generation growth. \textit{Second}, the invisibility of PV generation and native load can hinder designing optimal service restoration plans \cite{PV_handbook, distribution_hand_book}. During the restoration stage after an outage, the native demand might be several times higher than the pre-outage demand due to the simultaneous restarting of a large number of air-conditioning appliances. This anomalous demand should be estimated for optimal restoration plans because it can damage electric devices when simultaneously restoring a large number of customers. In practice, utilities multiply the normal native demand before outage by a ratio to estimate the anomalous demand during restoration. Also, utilities typically do not consider PVs as reliable restoration sources \cite{PV_handbook}. Therefore, separating normal native demand and generation is needed for estimating the restoration demand. \textit{Third}, the unobservability of native demand and solar generation might cause inaccurate reliability analysis. When evaluating a transmission system's reliability, each distribution system is generally simplified as a bus whose native load duration curve is constructed \cite{7078837,planning_guidebook}. For those utilities with a high-penetration PV integration, directly using the net demand to construct the load duration curve can significantly underestimate the actual native load \cite{rui_cheng_1}. This is because the net demand is typically smaller than the native demand due to the existence of PV generation. In contrast, using the native demand separated from the net demand can help construct more accurate load duration curves. In summary, disaggregating BTM PV generation and native demand from the recorded net demand can enhance distribution system observability and awareness and can also provide more accurate information for transmission system reliability analysis. 

Previous works on BTM PV generation disaggregation can be categorized into two types: \textit{Type I - Model-based approaches:} PV array performance model is employed to represent physical PV arrays. In \cite{SunDance}, a PV model is combined with a clear sky model to estimate customer-level solar generation. In \cite{Yi_Wang}, a virtual equivalent PV station model is utilized to represent the aggregate generation of BTM PVs within a region. In  \cite{YC_Zhang} and \cite{nan_peng_yu_2}, a physical PV model and a statistical model are utilized to estimate BTM solar generation and native demand, respectively. One primary disadvantage of these model-based approaches is that detailed PV array parameters or accurate meteorological data are required. However, in practice, these parameters are typically unavailable to utilities. Also, acquiring meteorological data might cause additional costs to utilities. \textit{Type II - Model-free approaches:} In \cite{kangping_li_solar} and \cite{ PV_cap_estimation}, net demands under heterogeneous weather conditions are employed to estimate BTM PV capacity, which is then multiplied by a standard solar power time series to infer BTM PV generations. In \cite{applied_energy}, native demand and PV generation are estimated using 1-second net demand data by identifying appliances' states, which are then leveraged to estimate appliance demands and solar power. Based on the variation difference between load and solar power, in \cite{indu_informatics}, an approach is proposed for estimating service transformer-level PV generation. In \cite{Hamid_Shaker}, regional-level generation is estimated by installing additional sensors to record typical PV generation profiles. In  \cite{CSSS_feeder}, feeder-level solar generation is estimated by utilizing net load measurements and a nearby PV farm’s generation readings. Using known native loads for customers without PVs and the generations for a limited number of observable PVs, in \cite{kangping_li_new}, the authors formulate an optimization process to estimate the aggregated native load and PV generation. In \cite{9548947}, a federated learning-based framework is proposed to probabilistically estimate community-level BTM solar generation. In \cite{Samuel_Talkington}, an approach is developed to estimate the reactive power by taking advantage of the correlation between the weekly nighttime and daytime native reactive power demands. Furthermore, previously in \cite{Fankun_Bu} and \cite{Fankun_Bu_1}, we have proposed two approaches for estimating the unknown BTM generation using measured solar power exemplars. One primary shortcoming of the model-free approaches is that they rely on contextual information, i.e., recorded solar power exemplars or meteorological data, which might bring additional costs to utilities.

\begin{figure}
\centering
\includegraphics[width=1\linewidth]{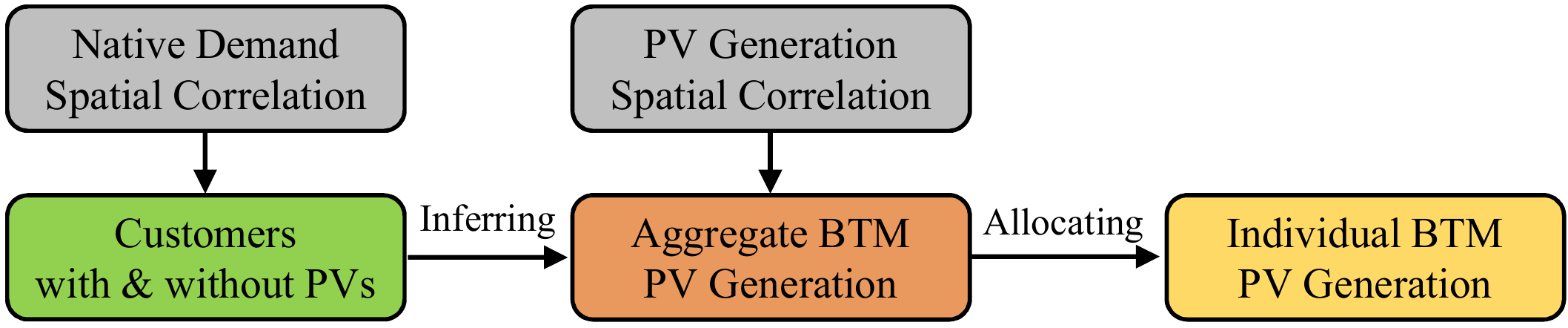}
\caption{\hl{Overall structure of the proposed BTM PV generation estimation approach.}}
\label{fig:overall_structure}
\end{figure}

Considering the shortcomings of previous approaches, this paper proposes a novel BTM PV generation and native demand estimation framework which does not require \textit{previously} recorded solar power and meteorological measurements. Our approach is based on two findings from real data. \hl{The first finding is the spatial correlation of native load, i.e., the native demands of two sizeable residential customer groups are strongly correlated and have highly homogeneous shapes. The second finding is the spatial correlation of solar power generation, i.e., the generations for two PVs in a distribution system are significantly correlated and have highly similar profiles.

Our proposed approach contains \textbf{two} layers: (1) At the aggregate level, the total generation of all BTM PVs is estimated by leveraging our first finding. (2) At the customer level, utilizing our second finding, the estimated aggregate BTM PV generation is allocated to individual customers. Utilizing the two findings improves our approach's robustness against the customer-level load uncertainty} \cite{load_uncertainty}.\hl{ The second layer contains three steps: first, our approach trains a model to produce multiple candidate generation time series, using solar power data generated by a publicly available tool. Second, our approach determines the peak generation for each PV. } Finally, the allocating procedure is formulated as an optimization problem. The overall structure of our proposed approach is shown in Fig. \ref{fig:overall_structure}. This paper has verified our proposed approach using real hourly native demand and PV generation data \cite{data_source}.

\hl{Smart meters can record individual customers' demands at an interval of one hour or shorter. Such fine-grained temporal and spatial granularity can give us more details than traditional monthly bills. Many researchers have developed advanced approaches to mine useful information from smart meter data. For example,} \cite{smart_meter_data_1} \hl{utilizes smart meter measurement to perform state estimation for enhancing distribution system observability, }\cite{smart_meter_data_2} \hl{employs water consumption data recorded by smart water meters to train aggregate water demand forecasters,} \cite{smart_meter_data_3} \hl{utilizes high-resolution phasor measurement units' data to conduct false data detection, and data redundancy strengthening, }\cite{smart_meter_data_4} \hl{converts smart meter data into manageable load profiles via linearizing load patterns. Our proposed approach takes advantage of smart meter data's temporal and spatial granularity to perform BTM generation estimation. }

\hl{The main contributions of our paper are summarized as follows: (1) This paper proposes an approach that does not rely on PV array parameters, historical meteorological data, or pre-recorded generation exemplars. This independence can significantly improve the viability of our approach because acquiring the above three types of information can bring challenges or additional costs for utilities. (2) Our approach only relies on the net demands of customers with PVs and the native demands of customers without PVs for estimating the aggregate-level PV generation. These two types of demands - net and native - are typically available to utilities, making our approach significantly practical. (3) Our approach innovatively estimates individual PV-installed customers' peak generations by mining net demand data. The peak generations are then utilized to estimate individual PV-installed customers' BTM generation time series. }

Throughout the paper, vectors are denoted using bold italic letters, and matrices are represented as bold non-italic letters. In addition, we adopt the sign convention that the native demand consumed by customers and the power output from PVs are both positive.

The rest of the paper is organized as follows: Section \ref{sec:findings} introduces our first and second findings regarding spatial correlation of native demand/generation. Section \ref{sec:aggre_level} presents how we estimate the aggregate generation for customers with PVs. Section \ref{sec:indiv_level} presents the procedure of formulating and solving an optimization problem to allocate the estimated aggregate generation to individual PVs. In Section \ref{sec:case_study}, case studies are analyzed. Section \ref{sec:conclusion} concludes the paper.

%%**************************%%
%                            %
%          *********         %
%                            %
%% *************************%%
\section{Spatial Correlation of Native Demand/PV Generation}\label{sec:findings}
\subsection{Finding 1: Native Demand Spatial Correlation between Two Sizeable Groups}\label{sec:finding_1}
By examining real residential native demand data, we find that once the customer numbers for two groups reach a certain level, their native demands are highly correlated. This finding is leveraged for estimating the \textit{aggregate} native demand time series for customers with PVs.
\begin{figure}
\centering
\includegraphics[width=0.84\linewidth]{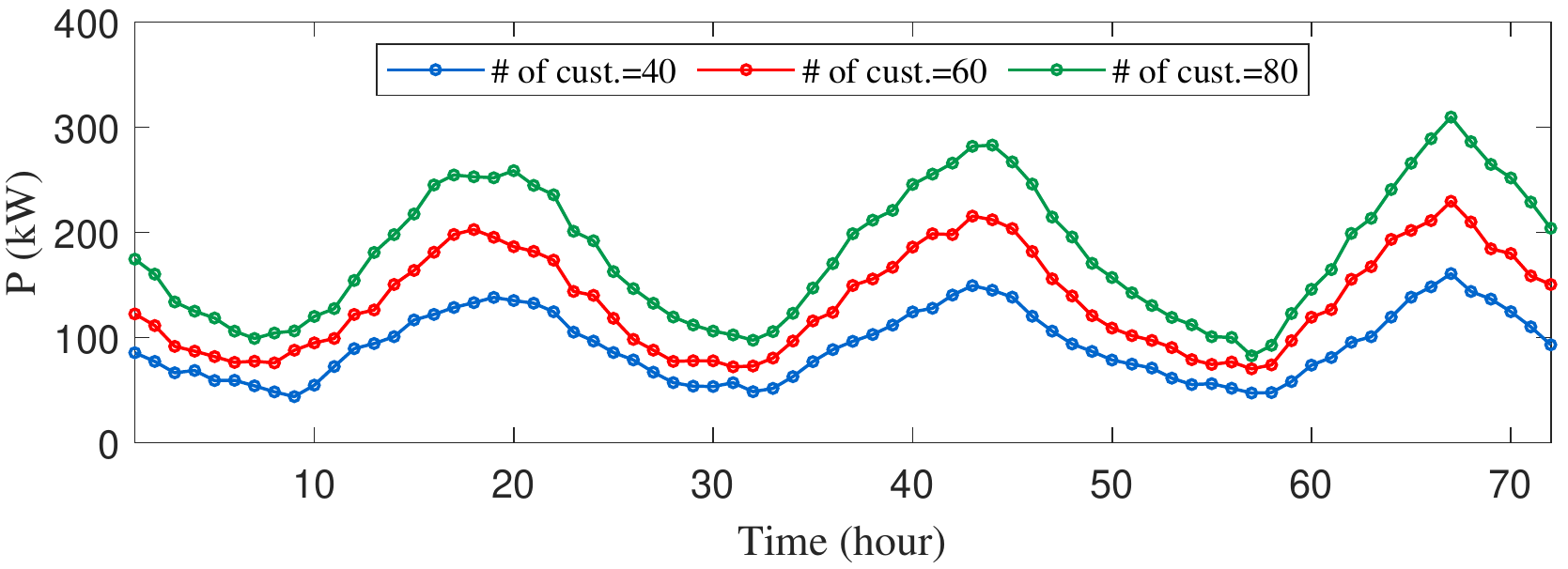}
\caption{Three-day actual native demand curves for three example groups with different customer numbers.}
\label{fig:native_load_curves}
\end{figure}

\begin{figure}
\centering
\includegraphics[width=0.84\linewidth]{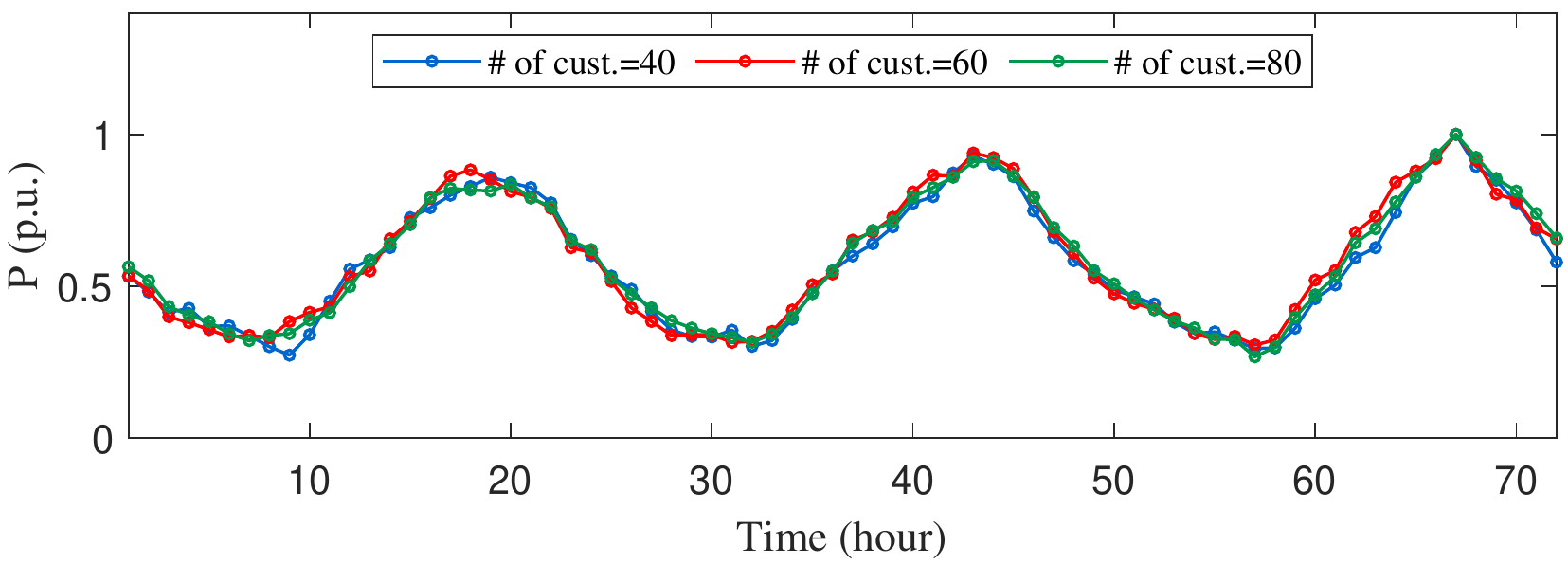}
\caption{Three-day normalized native demand curves for three example groups with different customer numbers.}
\label{fig:norm_native_load_curves}
\end{figure}

Specifically, we use native demand curves to illustrate the observed spatial correlation. Fig. \ref{fig:native_load_curves} presents real native demand curves for three example groups with different customer numbers, i.e., 40, 60, and 80, respectively. We can observe that these three curves demonstrate almost identical shapes, although they have different magnitudes. The high shape similarity can also be corroborated by Fig. \ref{fig:norm_native_load_curves}, which presents normalized native demand curves corresponding to the curves in Fig. \ref{fig:native_load_curves}. Note that the normalized curves are obtained by dividing the real curves in Fig. \ref{fig:native_load_curves} by their peaks, respectively. 

To stress the importance of Fig. \ref{fig:norm_native_load_curves}, we first define \textit{\textbf{two}} types of customer groups: the residential customers \textit{\textbf{with}} and \textit{\textbf{without}} PVs. These two customer groups are denoted as $C_w$ and $C_{o}$, respectively. For $C_o$, its native demand is recorded by smart meters. For $C_w$, we only know its net demand, and we do not know its native demand. Our goal is to estimate $C_w$'s unknown native demand and thus to estimate its PV generation. \hl{Therefore, Fig.} \ref{fig:norm_native_load_curves} \hl{inspires us that given the known native demand curve of $C_o$, we can infer the unknown native demand curve of $C_w$ by multiplying the native demand curve of $C_o$ by a \textit{ratio}, $r$.}

Since the native demands for the customers in $C_o$ are directly recorded by smart meters, the native demand curve of $C_o$ can be obtained by aggregating the native demand time series over the customers in $C_o$. The challenge for inferring the unknown native demand curve of $C_w$ is that the ratio, $r$, is unknown and needs to be estimated. \hl{The unknown of $r$ is caused by the unavailability of the native demand during the daytime for the customers in $C_w$. }This is because PV generates power during the daytime, which masks the native demand in the case of net metering. Thus, we cannot use daytime native demand to compute $r$. Instead, we use the nocturnal native demand to estimate $r$ because PV does not generate power during nighttime, and thus the nocturnal native demand for $C_w$ is known. Based on the above inference, we propose first utilizing the nocturnal native demand to compute a nocturnal native demand ratio, $r_n$, and then approximating $r$ as $r_n$. 

\begin{figure}
\centering
\includegraphics[width=0.45\linewidth]{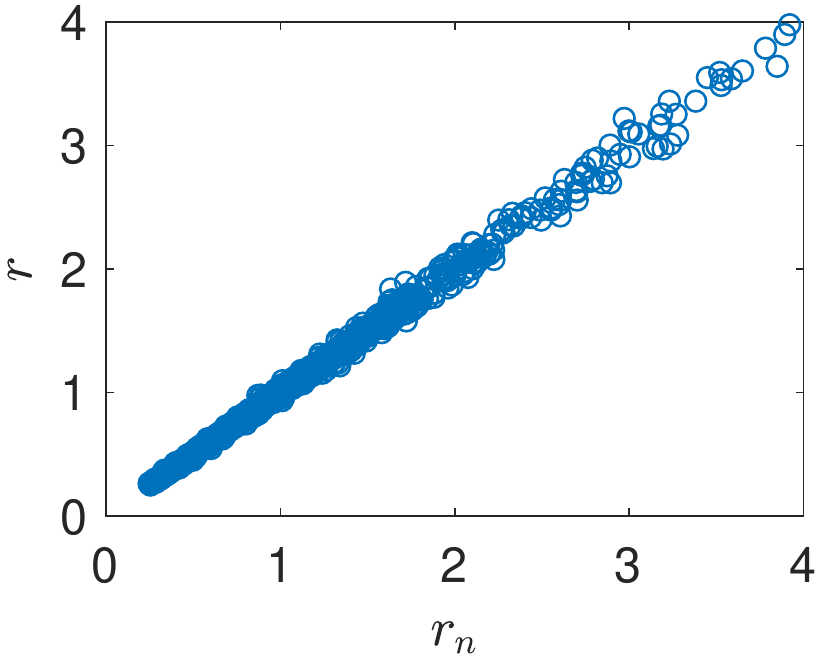}
\caption{The relationship between native demand ratio and the nocturnal native demand ratio between two example customer groups.}
\label{fig:demand_ratio}
\end{figure}

\hl{One \textit{pre-condition} for approximating $r$ as $r_n$ is that $r$ should be close to $r_n$.} To verify this condition, we randomly select two groups with different customer numbers ranging from 20 to 80. Then, for each group, the native demand time series are spatially aggregated over customers to obtain an aggregate native demand time series. After that, we compute $r$ using the two groups' native demand time series throughout a certain period, and compute $r_n$ using the two groups' native demand time series only during \textit{nighttime} within that period. Finally, we plot $r$ against $r_n$, as shown in Fig. \ref{fig:demand_ratio}. We can see that $r$ is almost identical with $r_n$. Therefore, we can accurately estimate $r$ by directly letting it equal $r_n$.

Once we obtain the estimate of $r$, we can compute the unknown native demand of $C_w$ by multiplying the known native demand of $C_o$ by the estimate of $r$. After that, estimating the unknown PV generation of $C_w$ is straightforward, i.e., by subtracting the recorded net demand measurements from the estimated native demand. 

\subsection{Finding 2: Generation Spatial Correlation between Two PVs}\label{sec:finding_2}
\begin{figure}
\centering
\includegraphics[width=0.808\linewidth]{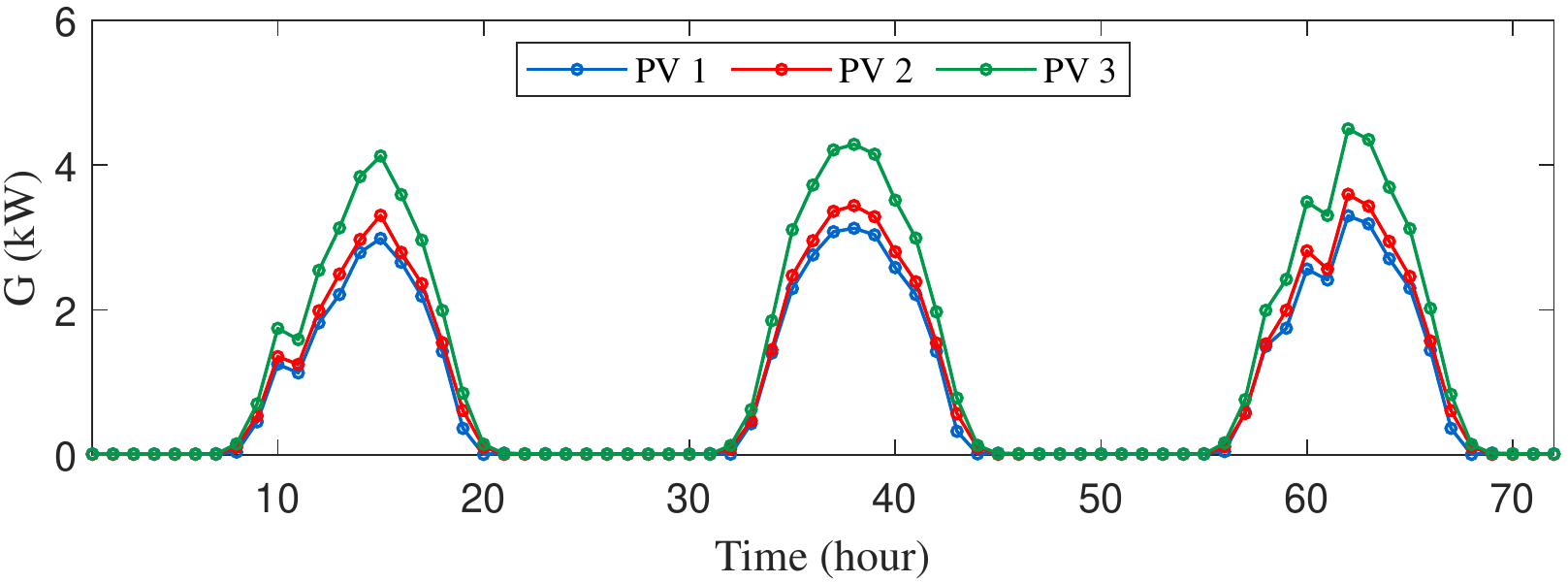}
\caption{Three-day real generation curves for three example PVs with different capacities.}
\label{fig:generation_curves}
\end{figure}

\begin{figure}
\centering
\includegraphics[width=0.808\linewidth]{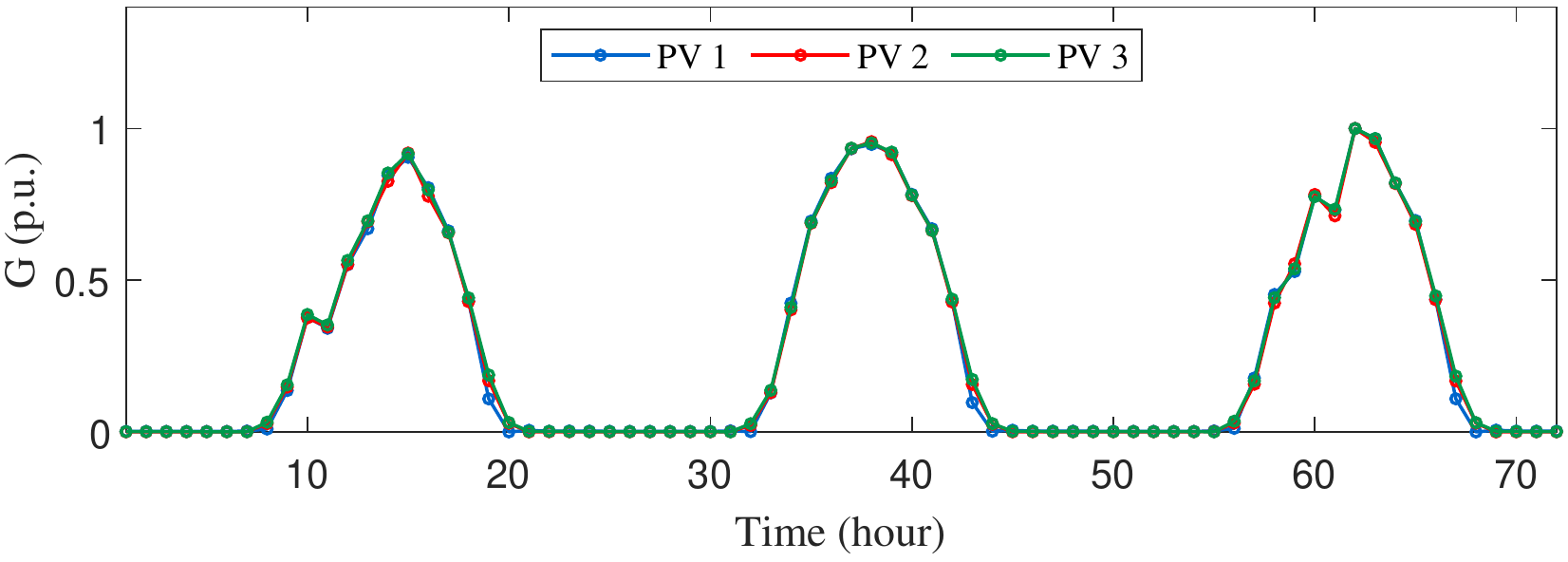}
\caption{Three-day normalized generation curves for three example PVs with different capacities.}
\label{fig:norm_generation_curves}
\end{figure}

There are two primary factors that determine the generation spatial correlation: (1) In most cases, a distribution system is geographically bounded in a small district. (2) The most widely available sampling resolution for smart meters is 1-hour. Under these two conditions, different PV arrays are subject to nearly identical meteorological inputs. Thus, the identical inputs can result in highly similar shapes among PV generation curves. Fig. \ref{fig:generation_curves} presents three example PV generation curves corresponding to different PV array capacities. Similar to the native demand curves for sizeable customer groups, these three generation curves also demonstrate significant spatial correlation, i.e., they possess highly similar shapes. This high similarity can also be corroborated by Fig. \ref{fig:norm_generation_curves}, where the normalized generation curves corresponding to the three curves in Fig. \ref{fig:generation_curves} overlap with each other. Most importantly, Fig. \ref{fig:generation_curves} and \ref{fig:norm_generation_curves} inspire us that estimating a BTM PV generation curve comes down to two steps: first, determine the generation curve's shape, and then determine its magnitude. This two-step method can notably simplify the estimation of unknown BTM PV generation time series. This is because compared to model-based methods, our approach is developed on the foundation of high similarity among generation curves; therefore, it requires significantly less information. 

%%**************************%%
%                            %
%              ******        %
%                            %
%% *************************%%
\section{Estimating Aggregate BTM PV Generation for Customers with PVs}\label{sec:aggre_level} 
\begin{figure*}
\centering
\includegraphics[width=0.75\linewidth]{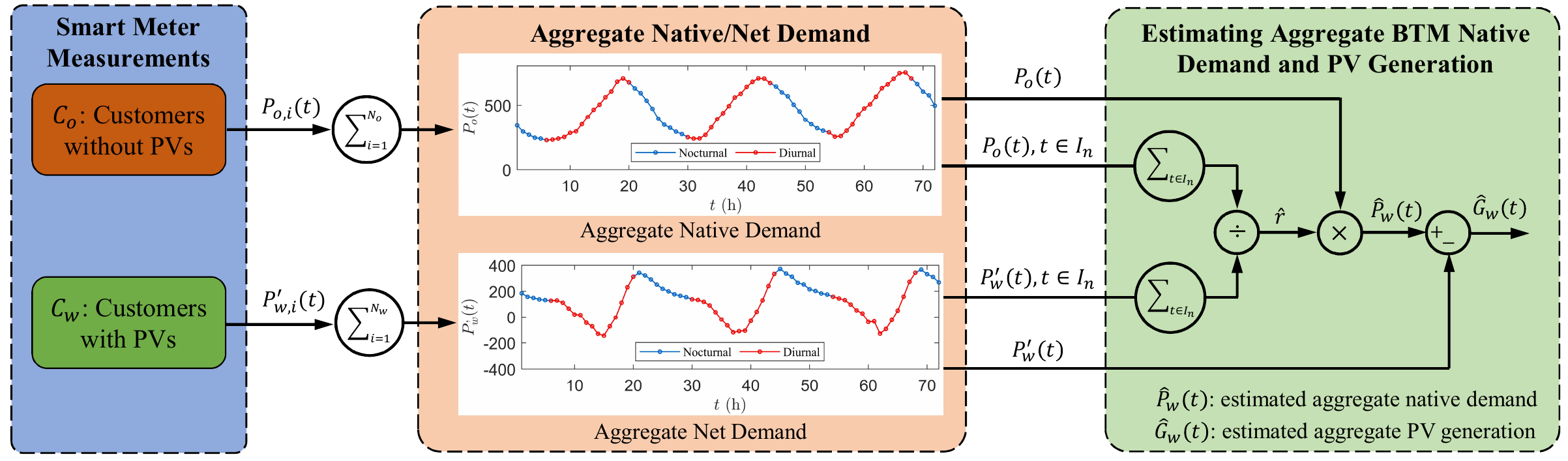}
\caption{Detailed structure of the proposed aggregate-level BTM PV generation/native demand estimation.}
\label{fig:aggregate_layer}
\end{figure*}

As elaborated in Section \ref{sec:finding_1},  the native demands of two sizeable customer groups are highly correlated. This high correlation inspires us that we can infer the unknown native demand of $C_w$ by multiplying the known native demand of $C_o$ by a ratio:
\begin{equation}  \label{eq:P_w}
\hat{\pmb{P}}_w = \pmb{P}_o r,
\end{equation}
where, $\hat{\pmb{P}}_w=\{\hat{P}_w(t)\}$ and $\pmb{P}_o=\{P_o(t)\}$, $t=1,...,T$, denote the estimated native demand time series for $C_w$ and the actual native demand time series for $C_o$, respectively. $T$ is the total number of native demands in a selected window (e.g., one month). $P_o(t)$ is computed by aggregating the measured native demands over customers without PVs:
\begin{equation}  \label{eq:P_o_t}
P_o(t) = \sum_{i=1}^{N_o}P_{o,i}(t), \quad t=1,...,T,
\end{equation}
where, $N_o$ represents the total number of customers in $C_o$, i.e., customers without PVs. $P_{o,i}(t)$ denotes the measured \textit{native} demand at time $t$ for the $i$'th customer in $C_o$.

In (\ref{eq:P_w}), $r$ denotes the native demand ratio between $C_w$ and $C_o$, and is defined as follows:
\begin{equation}  \label{eq:hat_r}
r = \frac{\Sigma_{t=1}^{T}P_w(t)}{\Sigma_{t=1}^{T}P_o(t)}.
\end{equation}
However, as presented in Section \ref{sec:finding_2}, since the \textit{diurnal} native demand for $C_w$ is masked by PV generation and unavailable to utilities, we need to estimate $r$ using \textit{nocturnal} native demand measurements. This approximation method is based on the observation that PV does not generate power during nighttime and the verification that $r$ and $r_n$ are almost identical. Specifically, we use $r_n$ to approximate $r$:
\begin{equation}  \label{eq:hat_r_n}
\hat{r} = r_n = \frac{\Sigma_{t \in I_{n}}^{}P_w(t)}{\Sigma_{t \in I_{n}}^{}P_o(t)},
\end{equation}
where, $I_{n}$ denotes the set of nighttime hours. In our paper, $I_n$ refers to the hours between 9:00 P.M. and 5:00 A.M. Note that for the hours in $I_n$, since PV does not generate power, $P_w(t)$ equals the known aggregate \textit{net} demand, $P_w'(t)$. Therefore,
\begin{equation}  \label{eq:hat_r_n_1}
\hat{r} = \frac{\Sigma_{t \in I_{n}}^{}P_w'(t)}{\Sigma_{t \in I_{n}}^{}P_o(t)},
\end{equation}
where, $P_w'(t)$ is computed by aggregating the measured net demands over customers in $C_w$:
\begin{equation}  \label{eq:P_w_prime_t}
P_w'(t) = \sum_{i=1}^{N_w}P_{w,i}'(t), \quad t=1,...,T,
\end{equation}
where, $N_w$ represents the total number of customers in $C_w$. $P_{w,i}'(t)$ denotes the measured \textit{net} demand at time $t$ for the $i$'th customer in $C_w$. 

Then, using the estimate of $r$ and the known native demand time series for $C_o$, we can apply (\ref{eq:P_w}) to compute the estimated native demand time series for $C_w$. Finally, inferring the PV generation time series for $C_w$, $\hat{\pmb{G}}_w = \{\hat{G}_w(t)\}$, $t=1,...,T$, is straightforward:
\begin{equation}  \label{eq:G_w}
\hat{\pmb{G}}_w = \hat{\pmb{P}}_w -  \pmb{P}_w',
\end{equation}
where, $\pmb{P}_w' = \{P_w'(t)\}$, $t=1,...,T$, denotes the known net demand time series for $C_w$. 

The above procedure for estimating the aggregate-level PV generation and native demand for $C_w$ are illustrated in Fig. \ref{fig:aggregate_layer}.

%%**************************%%
%                            %
%             ***            %
%                            %
%% *************************%%
\section{Estimating BTM PV Generation for Each Individual PV}\label{sec:indiv_level}  
Knowing the aggregate BTM PV generation and native demand might not be sufficient for some applications \cite{qianzhi, rui_cheng}. For example, some demand response schemes require known customer-level native demand \cite{kangping_li_solar}. Therefore, estimating individual customers' BTM native demand and PV generation is of significance.

To achieve this goal, we propose an approach to allocate the estimated aggregate PV generation/native demand time series to individual customers with PVs. As discussed in Section \ref{sec:finding_2}, estimating an individual PV's generation curve boils down to determining the generation curve's shape and its magnitude. \hl{In this section, our approach has three steps to perform allocating: (Step-I): generate candidate generation curves for individual PVs; (Step-II): estimate the peak generation for each PV; and (Step-III): allocate the estimated aggregate PV generation time series to individual PVs by solving an optimization problem.}

\subsection{Generating Diverse Candidate Generation Curves for Individual PVs }\label{sec:indiv_level_shape}  
As discussed earlier, in a geographically bounded distribution system, two primary factors determining a generation curve are the magnitude and shape. \hl{This subsection aims to generate candidate generation curves for those non-south-facing PVs. First, we train a regression model using the data generated by PVWatts Calculator. Then, we feed the estimated generation curve of a south-facing PV into the trained model to infer the targeted candidate generation curves for those non-south-facing PVs.}

\begin{figure}
\centering
\includegraphics[width=0.75\linewidth]{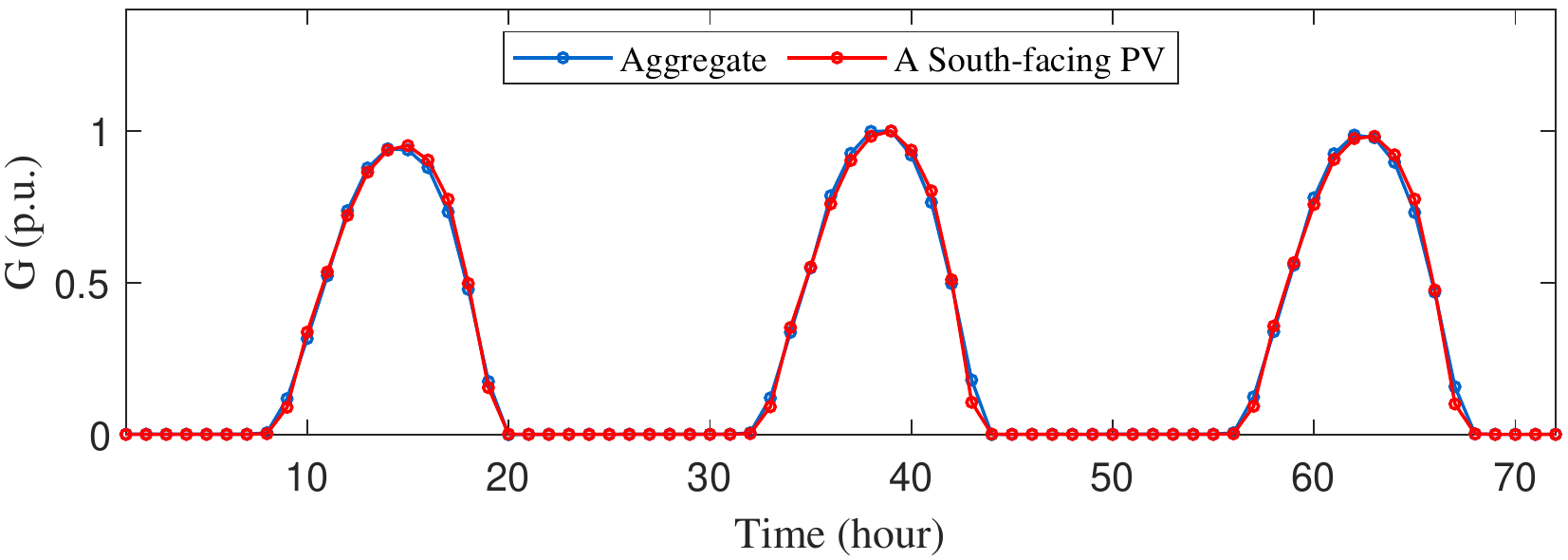}
\caption{Three-day normalized aggregate generation curve for all PVs and normalized generation curve for an individual PV facing south.}
\label{fig:nrm_agg_indi_gener_curves}
\end{figure}

In Section \ref{sec:aggre_level}, we have obtained the estimated time series for the aggregate generation of all PVs. One question is whether we can use that shape to represent the unknown shapes of individual PVs. To answer this question, we have conducted a numerical experiment. First, we normalized the aggregate generation curve of all PVs by dividing the aggregate generation time series by its peak. Then, in the same way, we normalized the generation curve of an example PV facing \textit{south}. The two normalized curves are plotted in Fig. \ref{fig:nrm_agg_indi_gener_curves}. It can be seen that the normalized curve corresponding to the aggregate generation for all PVs is highly similar to the normalized curve for a south-facing PV. One primary reason for this similarity is that the majority of residential PVs face south because a south-facing PV can typically generate more power than PVs in other directions. Most importantly, Fig. \ref{fig:nrm_agg_indi_gener_curves} tells us that a south-facing PV's generation curve can be accurately represented by the normalized aggregate generation curve of all PVs. %This is because conditioned on the same PV azimuth of south, different PVs  other parameters (e.g., PV capacity, efficiency, and array tilt) have a primary impact on the \textit{magnitude} of PV generation curve, rather than the \textit{shape}. 

\begin{figure}
\centering
\includegraphics[width=0.87\linewidth]{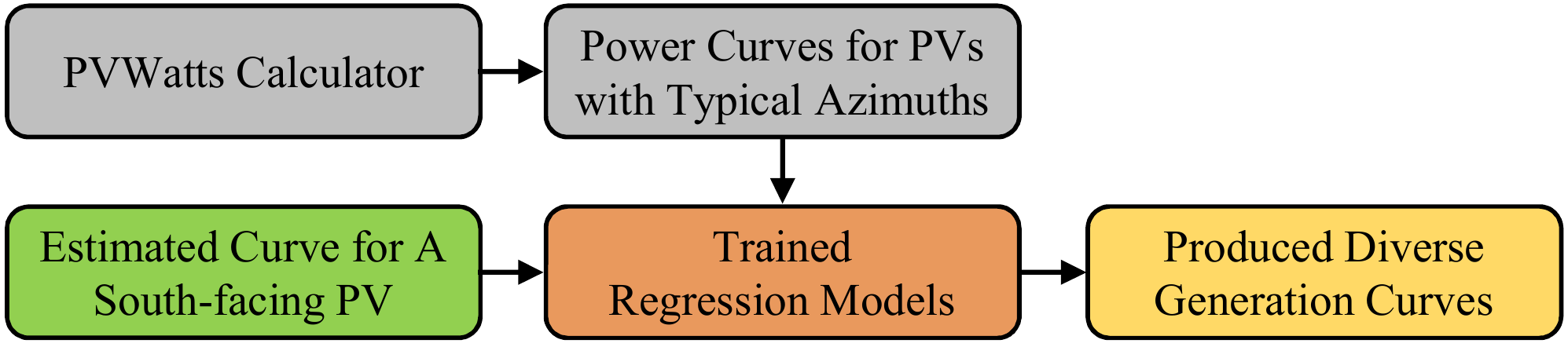}
\caption{Overall structure for producing diverse candidate PV generation curves using power output data generated by PVWatts Calculator.}
\label{fig:overall_framework_candidate_curves}
\end{figure}

Note that in distribution systems, in addition to the majority of south-facing PVs, there exist some residential PVs with other azimuths, such as east or west. These non-south-facing PVs' generation curves cannot be fully represented by the normalized aggregate PV generation curve in Fig. \ref{fig:nrm_agg_indi_gener_curves}. Specifically, compared to the normalized aggregate PV generation curve, the normalized generation curves for an east-facing PV and a west-facing PV are somewhat ``left-skewed" and ``right-skewed", respectively, as shown in Fig. \ref{fig:nrm_agg_indi_gener_curves_1}. Therefore, it is necessary to obtain candidate shapes for those non-south-facing PVs' generation curves. To achieve this goal, our basic idea is first to feed PV power data generated by PVWatts Calculator into a regression model to capture the relationship between the generations for a south-facing PV and a non-south-facing PV. Then, the aggregate generation curve estimated in Section \ref{sec:aggre_level}, which can accurately represent a south-facing PV's generation curve, is fed into the trained regression model to produce diverse generation curves corresponding to non-south azimuths. The overall structure is shown in Fig. \ref{fig:overall_framework_candidate_curves}:

\begin{figure}
\centering
\includegraphics[width=0.75\linewidth]{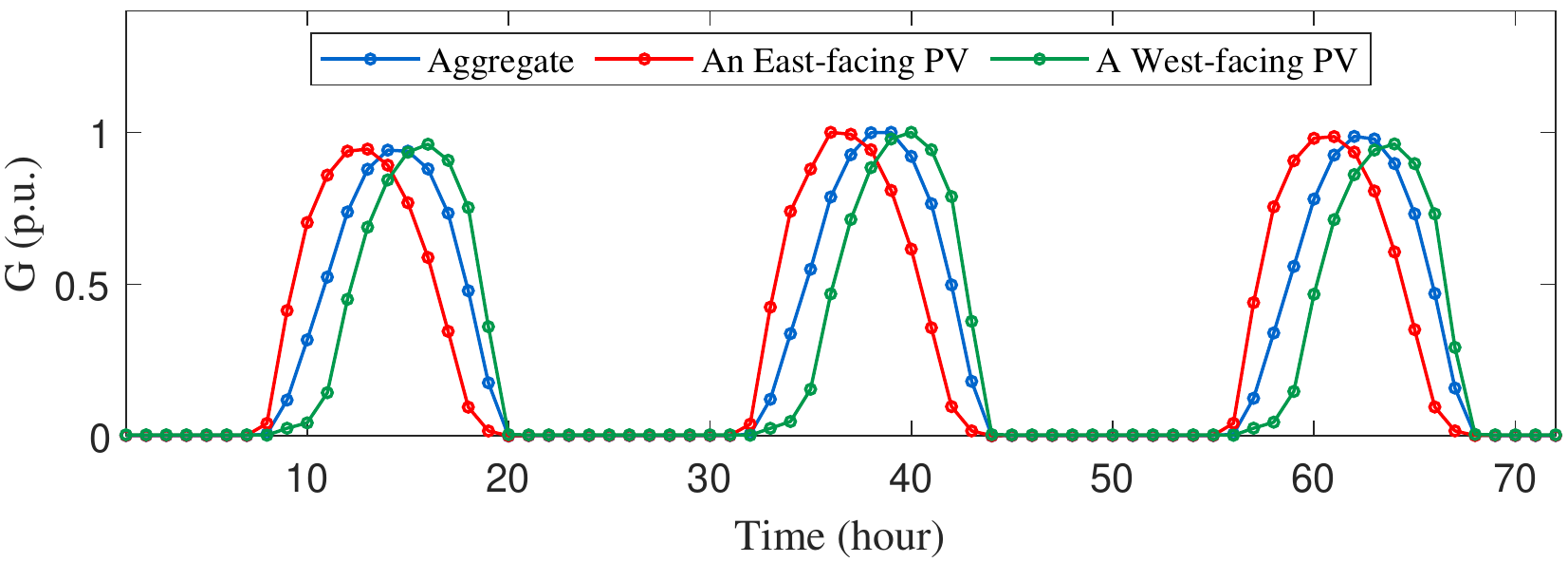}
\caption{Three-day normalized aggregate generation curve of all PVs and normalized generation curves for two example PVs facing east and west, respectively.}
\label{fig:nrm_agg_indi_gener_curves_1}
\end{figure}

\subsubsection{Training A Gaussian Process Regression Model}
Since the shape of a south-facing PV's generation curve can be approximated as the shape of the aggregate generation curve of all PVs, one intuitive way for inferring \textit{non-south-facing} PVs' candidate shapes is to produce diverse shapes based on the \textit{south-facing} PV's estimated generation curve. This idea is based on our observation that there exists a mapping between the generation curves for PVs with different azimuths. Therefore, one critical step for producing diverse candidate generation curves is to identify the relationship between a non-south-facing PV's generation curve and a south-facing PV's generation curve. %Then, using the identified relationship and the inferred normalized generation curve for a south-facing PV, we can infer the normalized generation curves for those non-south-facing PVs. 
To capture the relationship, first, we use PVWatts Calculator \cite{PV_watts}, an online application developed by the National Renewable Energy Laboratory (NREL), to generate power output data for PVs with typical azimuths, e.g., east, south, and west. Then, using the generated PV output power data, we train a Gaussian Process Regression (GPR) model to capture the relationship between the generation curve corresponding to a typical azimuth except for south (e.g., east) and the generation curve corresponding to the azimuth of the south. %Once the relationship has been captured, the estimated normalized aggregate generation curve is passed to the trained GPR model to produce normalized generation curves corresponding to the typical azimuth except for the south. 
The primary reason for selecting GPR is that after running numerical tests, GPR demonstrated a relatively better performance when applied to our dataset than some other state-of-the-art nonlinear regression models, such as the Support Vector Machine model and the Polynomial regression model. 

Specifically, first, we use PVWatts Calculator to generate time-series data for a south-facing PV and a PV with other typical azimuth (e.g., east). Then, each time series is normalized so that the peak generation is 1 p.u. The two normalized time series corresponding to the south-facing PV and the non-south-facing PV are denoted as $\pmb{G}_s^*=\{G_s^*(t)\}$ and $\pmb{G}_{ns}^*=\{G_{ns}^*(t)\}$, $t=1,...,T$, respectively. $G_s^*(t)$ and $G_{ns}^*(t)$ denote the normalized generation at time $t$ for a south-facing PV and a non-south-facing PV, respectively. Our goal is to use $G_{s}^*(t)$ to explain $G_{ns}^*(t)$ because PVs in a geographically bounded distribution system typically have highly correlated generations. By conducting numerical experiments, we find that in addition to $G_s^*(t)$, the hour-in-day, $H_d(t)$, and day-in-year, $D_y(t)$, are also related with $G_{ns}^*(t)$. Therefore, we use ${G}_{s}^*(t)$, $H_d(t)$, and $D_y(t)$ as the input variables and ${G}_{ns}^*(t)$ as the output variable, respectively, to train a GPR model. The function of GPR is to capture the relationship between $G_{ns}^*(t)$ and $G_{s}^*(t)$. The basic idea behind GPR is that if the distance between two explanatory variables is small, the difference between their corresponding dependent variables will also be relatively small. Specifically, the output, $G_{ns}^*(t)$, is denoted as a function of the input vector, $\pmb{X}^*(t)$:
\begin{equation}  \label{eq:f_function}
G_{ns}^*(t) = f(\pmb{X}^*(t)),
\end{equation}
where, $\pmb{X}^*(t)=[G_{s}^*(t), H_d(t),D_y(t)]^\mathsf{T}$. For GPR, $f(\pmb{X}^*(t))$ is assumed to be a random variable reflecting the uncertainty of functions evaluated at $\pmb{X}^*(t)$. Specifically, the function $f(\pmb{X}^*(t))$ is distributed as a Gaussian process:
\begin{equation}  \label{eq:GPR_definition}
f \big(\pmb{X}^*(t) \big) \sim \mathcal{GP} \big(\mu(\pmb{X}^*(t)), K(\pmb{X}^*(t), \pmb{X}^*(t')) \big),
\end{equation}
where, $\mu(\pmb{X}^*(t))$ represents the expected value of $f(\pmb{X}^*(t))$, i.e., the value of $G_{ns}^*(t)$. The covariance function, $K(\pmb{X}^*(t), \pmb{X}^*(t'))$, represents the dependence between $G_{ns}^*(t)$'s at different times. In our problem, the covariance function, $K(\cdot,\cdot)$, is specified by the Squared Exponential Kernel function expressed as:
\begin{equation}  \label{eq:sqrt_exp_kernel}
K \big(\pmb{X}^*(t),\pmb{X}^*(t') \big) = \sigma_f^2 \text{exp} \bigg(-\frac{||\pmb{X}^*(t) - \pmb{X}^*(t')||_2^2}{2 \sigma^2}\bigg),
\end{equation}
where, $||\cdot||_2$ represents $l_2$-norm, $\sigma_f$ and $\sigma$ are hyper-parameters, which are determined using cross-validation. Intuitively, (\ref{eq:sqrt_exp_kernel}) measures the distance between $\pmb{X}^*(t)$ and $\pmb{X}^*(t')$, which can also reflect the similarity between $G_{ns}^*(t)$ and $G_{ns}^*(t')$.

Note that $G_{s}^*(t)$ and $G_{ns}^*(t)$ are generated solar powers using PVWatts Calculator; thus, they are known and a $T$-dimensional joint Gaussian distribution can be constructed as:
\begin{equation}   \label{eq:N_dim_Gau} 
\left[
\begin{array}{c}
f \big(\pmb{X}^*(1) \big)  \\
\vdots \\
f \big(\pmb{X}^*(T) \big)
\end{array}
\right] 
\sim \mathcal{N}
\Big(
\pmb{\mu}^*, \pmb{\Sigma}^*
\Big),
\end{equation}
where, 
\begin{subequations}  \label{eq:mu_sigma}
\begin{equation}
\pmb{\mu}^*=
\left[
\begin{array}{c}
\mu \big(\pmb{X}^*(1) \big)  \\
\vdots \\
\mu \big(\pmb{X}^*(T) \big)
\end{array}
\right],
\end{equation}    
\begin{equation}
\pmb{\Sigma}^*=
\left[
\begin{array}{ccc}
K\big(\pmb{X}^*(1),\pmb{X}^*(1) \big) & \cdots  & K\big(\pmb{X}^*(1),\pmb{X}^*(T) \big) \\
\vdots  & \ddots & \vdots\\
K\big(\pmb{X}^*(T),\pmb{X}^*(1) \big) & \cdots  &  K\big(\pmb{X}^*(T),\pmb{X}^*(T) \big)
\end{array}
\right].
\end{equation}
\end{subequations}

The joint Gaussian distribution formulated in (\ref{eq:N_dim_Gau}) represents a trained non-parametric model, which captures the relationship between $G_{ns}^*(t)$ and $G_{s}^*(t)$.

\subsubsection{Inferring A Non-south-facing PV's Generation Curve}
As shown in Fig. \ref{fig:nrm_agg_indi_gener_curves}, the normalized generation curve for a south-facing PV, $\pmb{G}_{s}=\{G_{s}(t)\}$, $t=1,...,T$, can be approximated as the normalized estimated aggregate generation curve for all PVs:
\begin{equation}  \label{eq:norm_G_south}
\pmb{G}_{s}  =  \frac{\hat{\pmb{G}}_w}{\hat{G}_m},
\end{equation}
where, $\hat{G}_m$ denotes the peak of $\hat{\pmb{G}}_w$. To infer the unknown generation time series for a non-south-facing PV, $\pmb{G}_{ns}=\{G_{ns}(t)\}$, $t=1,...,T$, we assume $G_{ns}(t)$ is a function of $G_{s}(t)$, i.e., $G_{ns}(t)=f(G_{s}(t))$. By appending $f(G_{s}(t))$ to the end of (\ref{eq:N_dim_Gau}), an ($N+1$)-dimensional joint Gaussian distribution can be constructed as:
\begin{multline}\label{eq:N_k_dim_Gau_1} 
\quad \quad \quad \quad 
\left[
\begin{array}{c}
G_{ns}^*(1)  \\
\vdots \\
G_{ns}^*(T) \\
G_{ns}(t)
\end{array}
\right] 
=
\left[
\begin{array}{c}
f(\pmb{X}^*(1))  \\
\vdots \\
f(\pmb{X}^*(T)) \\
f(\pmb{X}(t))
\end{array}
\right] \\
\sim \mathcal{N}
\Big(
\left[
\begin{array}{c}
\pmb{\mu}_*\\
{\mu}_1
\end{array}
\right],
\left[
\begin{array}{cc}
\pmb{\Sigma}^{*} & \pmb{\Sigma}_{*1}\\
\pmb{\Sigma}_{*1}^T & {\Sigma}_{11}
\end{array}
\right]
\Big),
\quad \quad \quad 
\end{multline}   
where, $\pmb{X}(t)=[G_s(t), H_d(t), D_y(t)]^\mathsf{T}$ is a vector of explanatory variables. $\pmb{\Sigma}_{*1}$ represents the training-test set covariances and ${\Sigma}_{11}$ is the test set covariance. Since $G_{ns}^*(t)$, $\pmb{X}^*(t)$, and $\pmb{X}(t)$ are known, using the Bayes rule, the distribution of $G_{ns}(t)$ conditioned on $\pmb{G}_{ns}^*$ can be computed as follows:
\begin{equation}  \label{eq:inferred_non_sou_fac_G}
G_{ns}(t) | \pmb{G}_{ns}^* \sim \mathcal{N} ({\mu}_1(t), \Sigma_1(t)),
\end{equation}
where, ${\mu}_1(t)=\pmb{\Sigma}_{*1}^T{\pmb{\Sigma}^*}^{-1} \pmb{G}_{ns}^*$ and $\Sigma_1(t)=\Sigma_{11}-\pmb{\Sigma}_{*1}^T {\pmb{\Sigma}^*}^{-1}\pmb{\Sigma}_{*1}$. Note that ${\mu}_1(t)$ denotes the most probable value of the estimated generation at time $t$ for a non-south-facing PV. By conducting the above inferring procedure for all the $t$'s, we can obtain a candidate generation time series corresponding to a \textit{particular} typical PV azimuth. Since there are multiple typical azimuths, such as east, and west, we can infer multiple candidate PV generation time series:
\begin{equation}  \label{eq:mult_cand_curves}
\pmb{G}_{ns}^j=\{G_{ns}^j(t)\}, \quad t=1,..., T, \quad j=1,...,N_{ns},
\end{equation}
where, $G_{ns}^j(t)$ denotes the inferred PV generation at time $t$, for the $j$'th typical non-south-facing azimuth. $N_{ns}$ denotes the total number of typical non-south-facing PV azimuths and is determined by conducting numerical experiments.

\subsection{Estimating Peak Generation for Each Individual PV}\label{sec:indiv_level_peak} 
Simply knowing the candidate shapes for unknown generation curves is insufficient for allocating the estimated aggregate generation to individual PVs. As discussed earlier, we should also know the magnitudes for the candidate generation curves. To estimate the peak generation, we employ our observation from real data that the peak generation is almost identical with the difference between the minimum diurnal native demand and the minimum net demand. 
\begin{figure}
\centering
\includegraphics[width=0.64\linewidth]{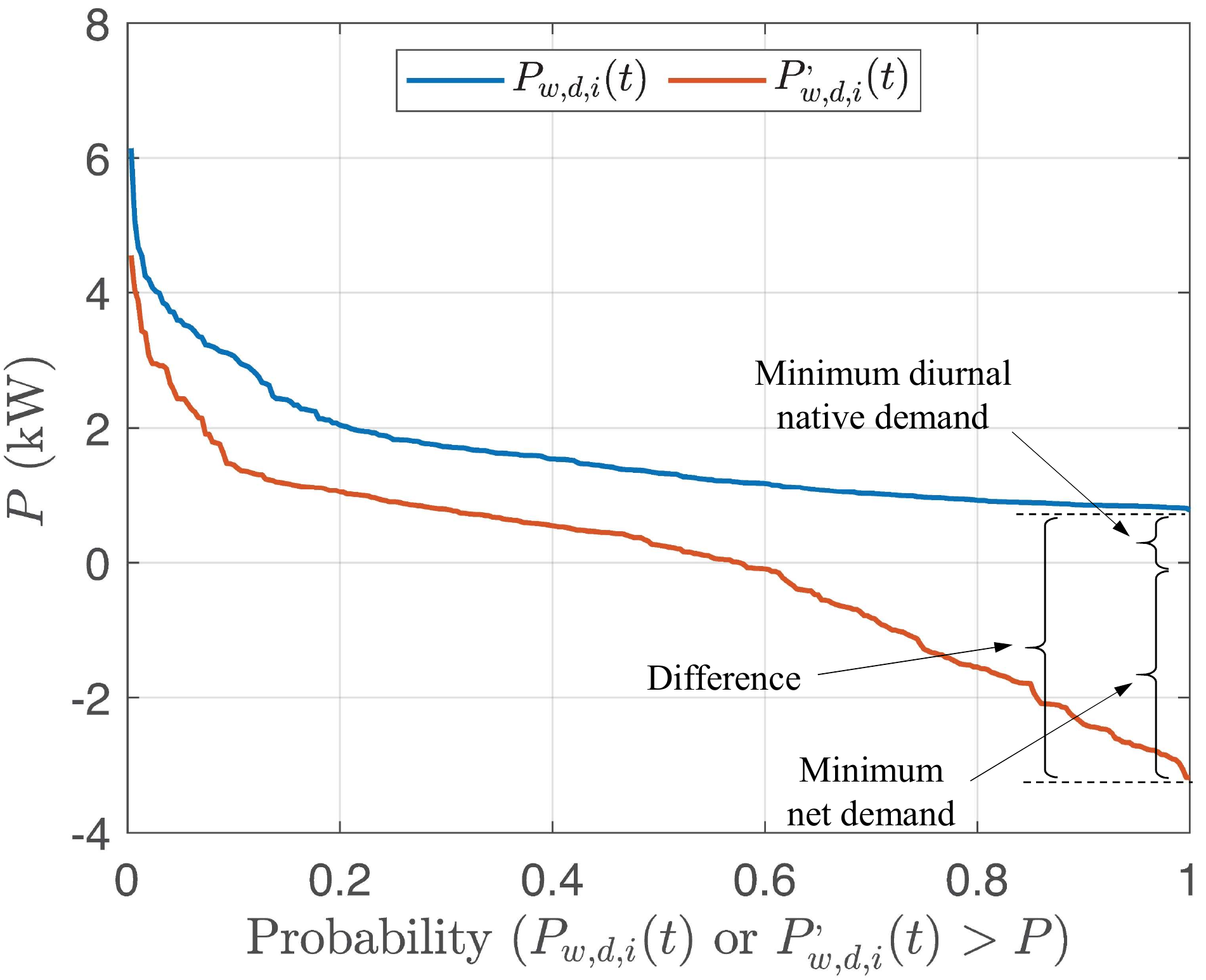}
\caption{Load duration curves for an example customer's diurnal native demand and diurnal net demand.}
\label{fig:duration_curves}
\end{figure}

\begin{figure}[htbp]
\centering
\subfloat[Spring\label{sfig:diff_vs_peak_G_spring}]{
\includegraphics[width=0.38\linewidth]{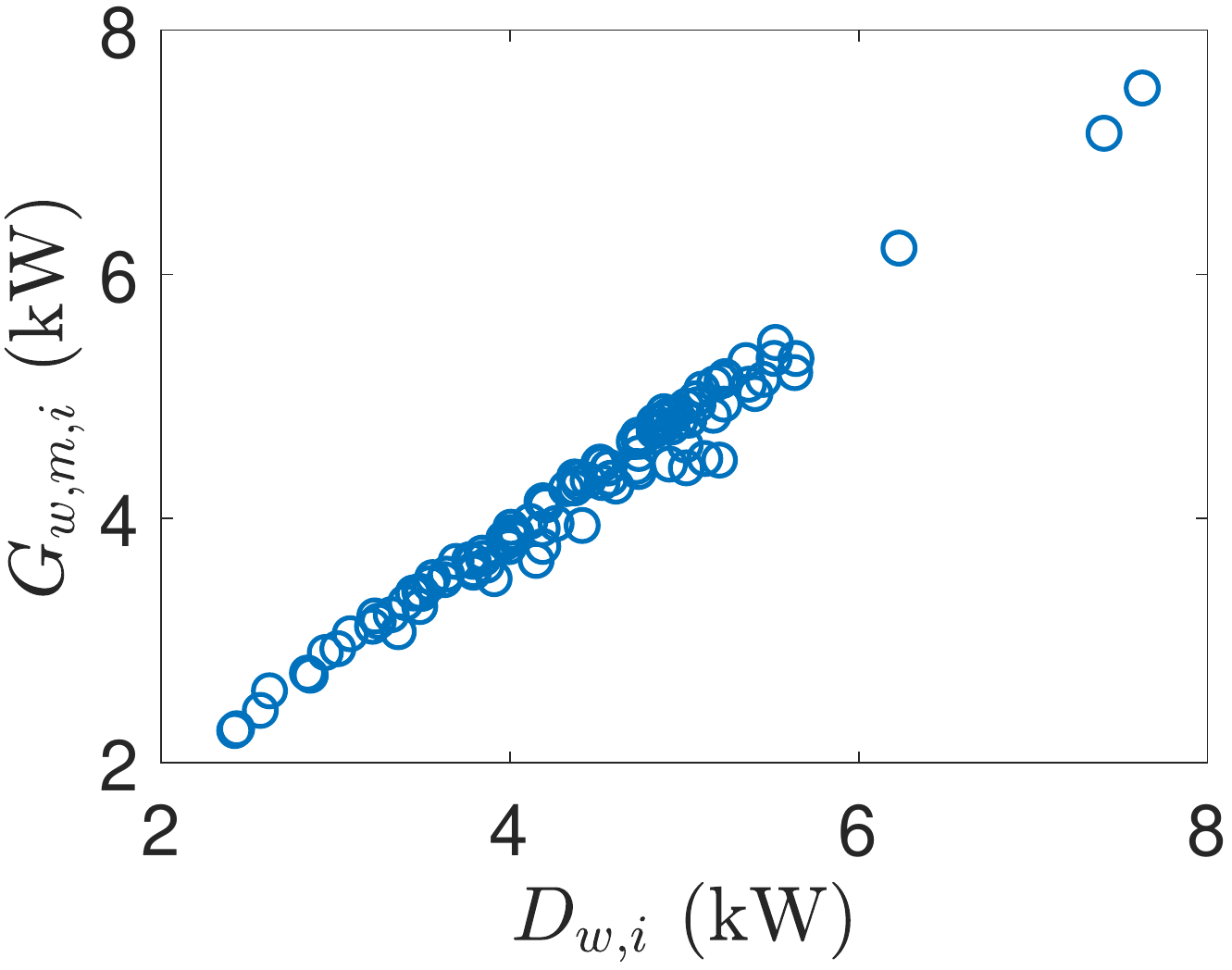}
}
\hfill
\subfloat[Summer\label{sfig:diff_vs_peak_G_summer}]{
\includegraphics[width=0.38\linewidth]{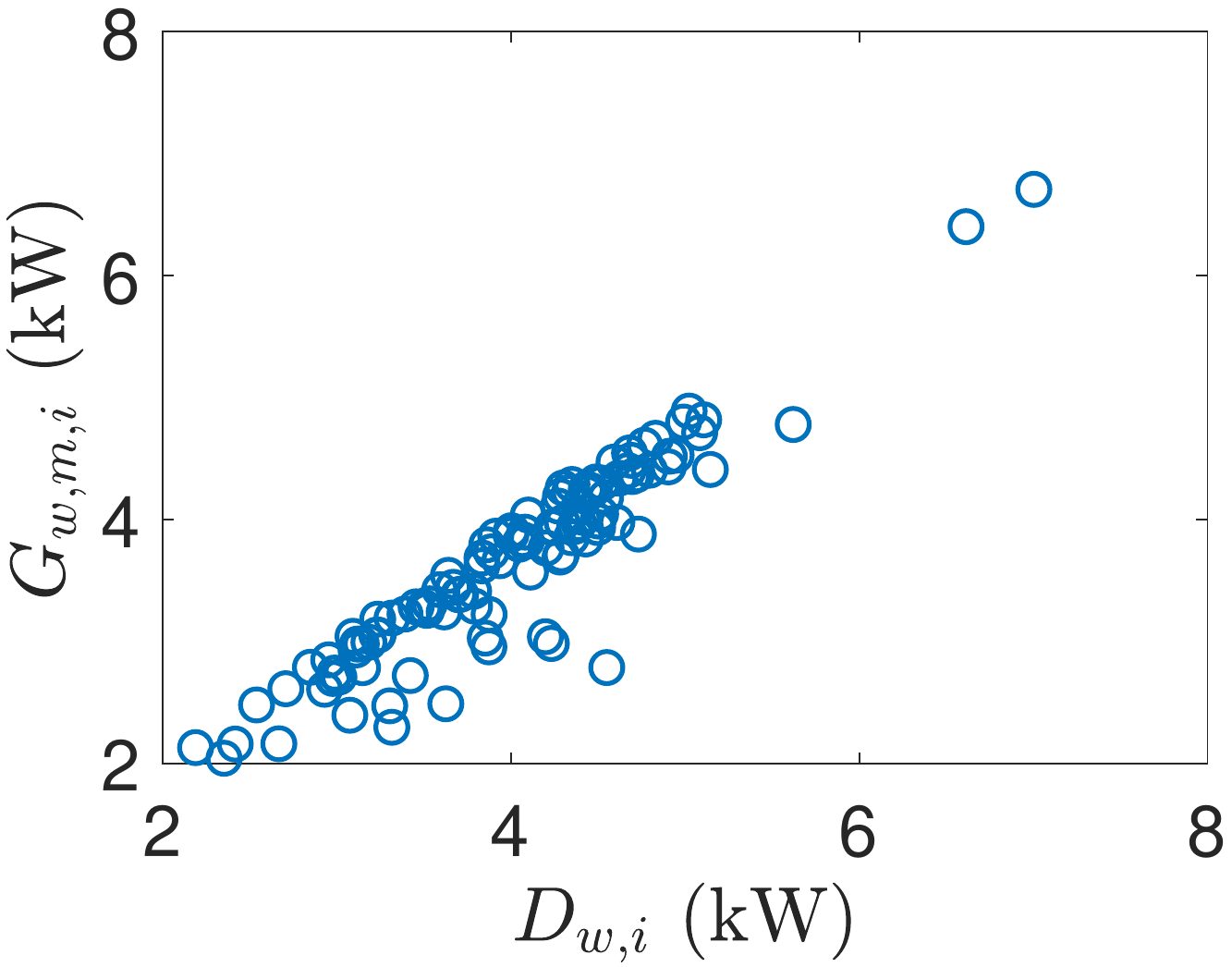}
}
\caption{The relationship between peak generation and the difference between minimum diurnal \textit{native} demand and minimum \textit{net} demand.}
\label{fig:diff_vs_peak_G}
\end{figure}

\begin{figure}[htbp]
\centering
\subfloat[Spring\label{sfig:noc_min_P_vs_diu_min_P_spring}]{
\includegraphics[width=0.44\linewidth]{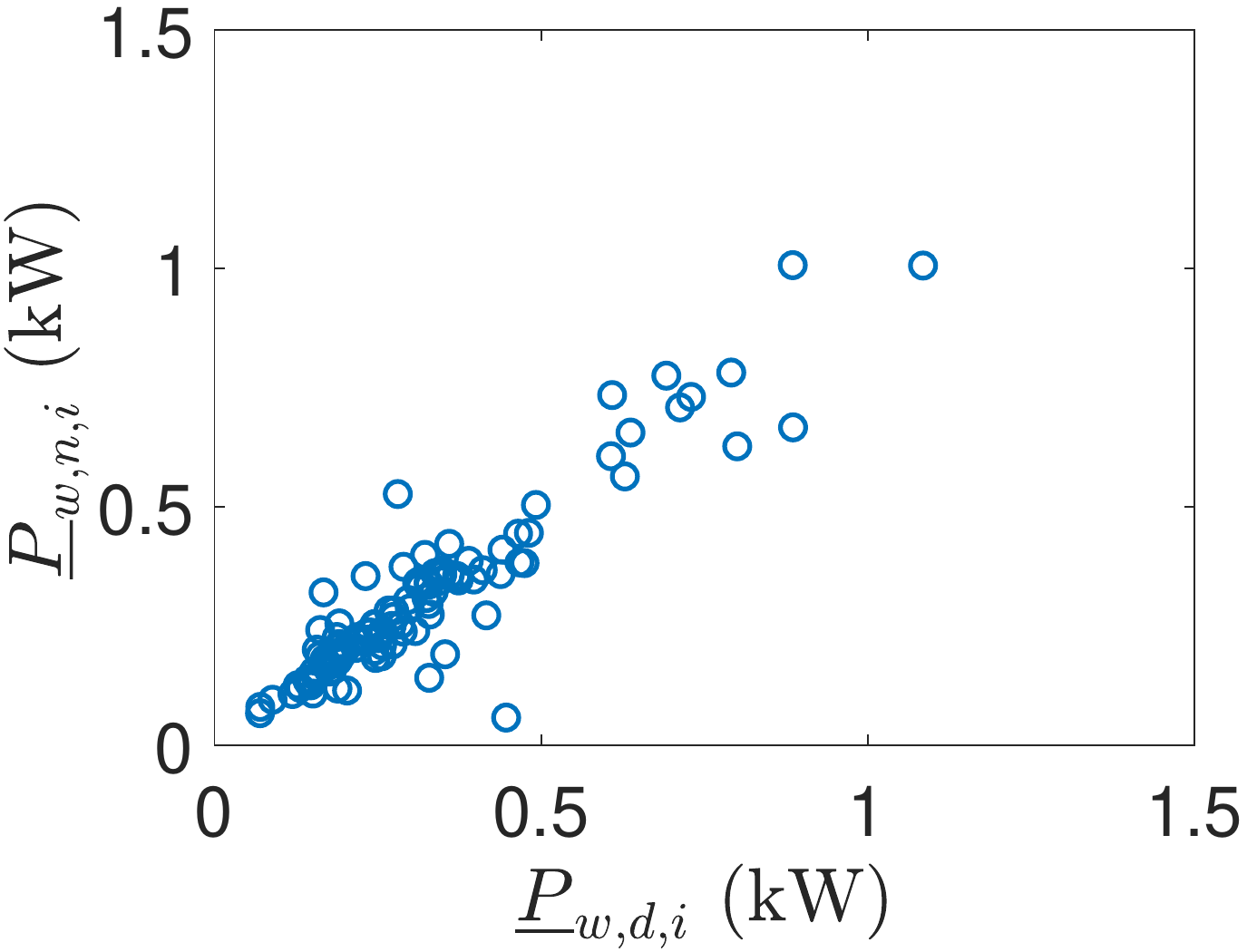}
}
\hfill
\subfloat[Summer\label{sfig:noc_min_P_vs_diu_min_P_summer}]{
\includegraphics[width=0.44\linewidth]{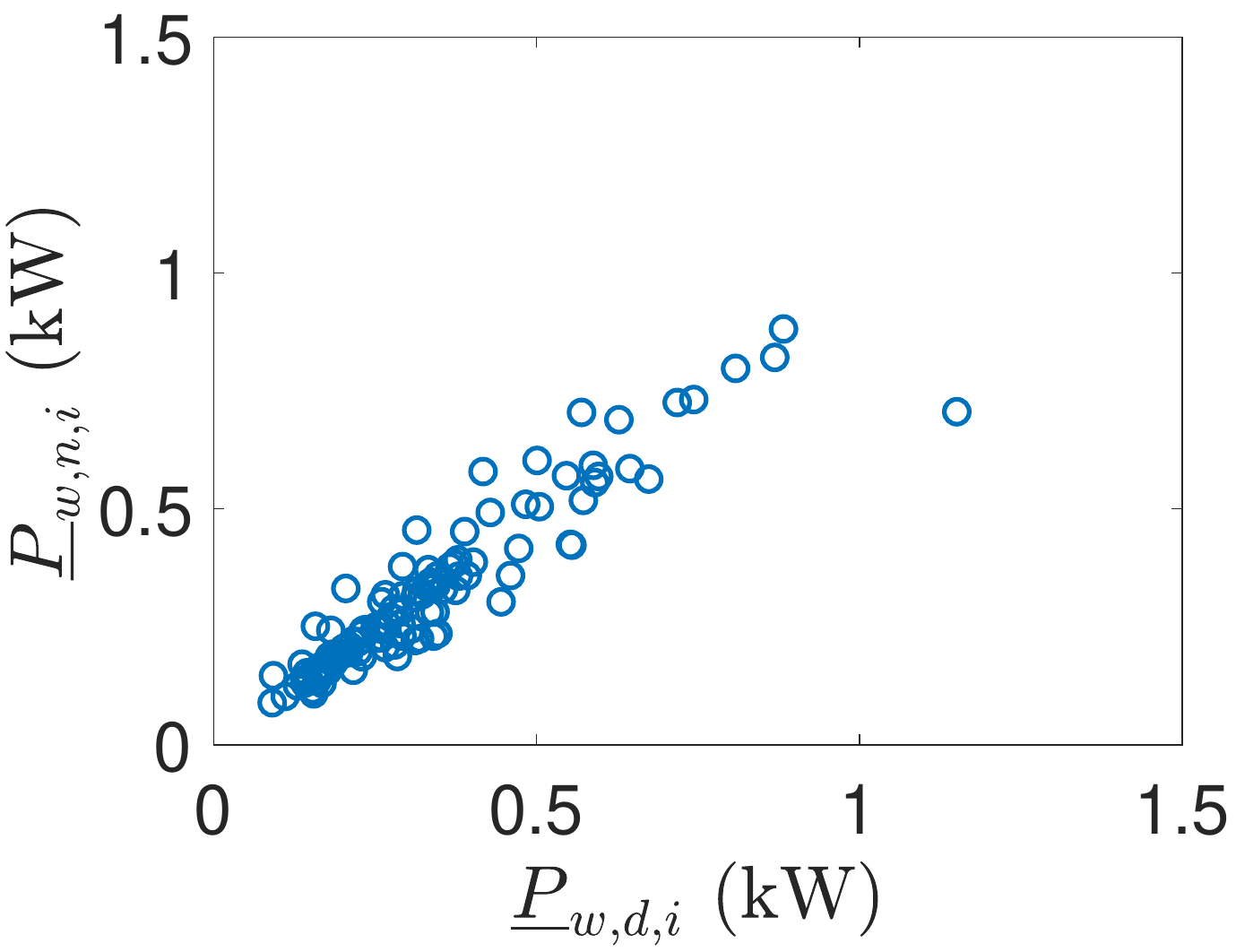}
}
\caption{The relationship between minimum \textit{diurnal} native demand and minimum \textit{nocturnal} native demand.}
\label{fig:noc_min_P_vs_diu_min_P}
\end{figure}

Specifically, to explain our observation regarding the correlation, we start with Fig. \ref{fig:duration_curves}, showing the load duration curves for the $i$'th customer's diurnal \textit{native} demand, $P_{w,d,i}(t)$, and diurnal \textit{net} demand, $P_{w,d,i}'(t)$. Thus, we can compute the difference between the minimums of $P_{w,d,i}(t)$ and $P_{w,d,i}'(t)$:
\begin{equation}  \label{eq:D_{w,i}}
D_{w,i} = \underbar{$P$}_{w,d,i} - \underbar{$P$}_{w,d,i}',
\end{equation}
where, $\underbar{$P$}_{w,d,i}$ and $\underbar{$P$}_{w,d,i}'$ denote the minimums of $P_{w,d,i}(t)$ and $P_{w,d,i}'(t)$ during a selected window, respectively. Note that $\underbar{$P$}_{w,d,i}$ is positive, and $\underbar{$P$}_{w,d,i}'$ is negative. Then, our finding is that $D_{w,i}$ is highly similar to the peak generation, $G_{w,m,i}$, as shown in Fig. \ref{fig:diff_vs_peak_G}. This relationship inspires us to approximate $G_{w,m,i}$ as $D_{w,i}$:
\begin{equation}  \label{eq:G_{w,m,i}}
\hat{G}_{w,m,i} = D_{w,i}, \quad i=1,..,N_w,
\end{equation}
where, $\hat{G}_{w,m,i}$ is the estimate of $G_{w,m,i}$. However, one challenge is that $D_{w,i}$ depends on $\underbar{$P$}_{w,d,i}$, which is unknown due to BTM PV generation. Therefore, we need to estimate $\underbar{$P$}_{w,d,i}$, which is involved with another finding from real native demand data. Specifically, as shown in Fig. \ref{fig:noc_min_P_vs_diu_min_P}, the minimum \textit{diurnal} native demand, $\underbar{$P$}_{w,d,i}$, can be approximated as the minimum \textit{nocturnal} native demand, $\underbar{$P$}_{w,n,i}$:
\begin{equation}  \label{eq:diu_min_P_i}
\underbar{$P$}_{w,d,i} \approx \underbar{$P$}_{w,n,i}, \quad i=1,..,N_w.
\end{equation}
Note that since PV does not generate power during nighttime, $\underbar{$P$}_{w,n,i}$ is known to utilities. Finally, using the estimate of $\underbar{$P$}_{w,d,i}$ and the known $\underbar{$P$}_{w,d,i}'$, we can compute $D_{w,i}$ using (\ref{eq:D_{w,i}}), and then compute $\hat{G}_{w,m,i}$ using (\ref{eq:G_{w,m,i}}).

\subsection{Allocating the Estimated Aggregate PV Generation to Individual PVs }\label{sec:optimization_formulation}
Sections \ref{sec:aggre_level}, \ref{sec:indiv_level_shape}, and \ref{sec:indiv_level_peak} provide the estimated aggregate generation time series of all PVs, inferred candidate generation curves for individual PVs, and estimated generation peaks for individual PVs, respectively. Therefore, estimating individual PVs' generation curves comes down to allocating the estimated aggregate generation time series to individual PVs. This allocating procedure is formulated as an optimization process:
\begin{subequations}  \label{eq:overall}
\begin{flalign}  \label{eq:overall_a}
\underset{\mathbf{K}, \pmb{\gamma}}{\textit{min}} \;
\, || \mathbf{G}_e * \mathbf{K} * \pmb{1} - \hat{\pmb{G}}_w ||_2^2 + \lambda * ||\pmb{\gamma}||_2^2
\end{flalign}
\vspace{-14pt} 
\begin{flalign}  \label{eq:overall_b}
\quad \quad \quad \quad \quad   \textit{s.t.} \; \;
  \mathbf{G}_e * \mathbf{K} \le \pmb{1}* (\hat{\pmb{G}}_{w,m} + \pmb{\gamma})^\mathsf{T},  &&
\end{flalign}
\vspace{-14pt}  
\begin{flalign}  \label{eq:overall_d}
 \quad\quad\quad   \quad \quad \quad \quad 
 \pmb{0} \le \pmb{\gamma} \le P_0 * \pmb{1},   &&
\end{flalign}
\end{subequations}
where,  $\mathbf{G}_e=[\pmb{G}_s, \pmb{G}_{ns}^1, ..., \pmb{G}_{ns}^{N_{ns}}]$ is a $T$-by-$N_e$ matrix, which denotes a collection of candidate generation curves. $N_e=N_s+1$ denotes the total number of candidate generation curves. $\mathbf{K}=[\pmb{K}_1,...,\pmb{K}_{N_w}]$ is an $N_e$-by-$N_w$ matrix of decision variables, which denote the weights assigned to candidate generation curves for individual PVs. $\pmb{K}_i,i=1,...,N_w$, is an $N_e$-by-1 vector, which denotes the weights assigned to candidate generation curves for the $i$'th PV. The first $\pmb{1}$ is an $N_w$-by-1 vector of ones. $\mathbf{G}_e * \mathbf{K}$ results in a $T$-by-$N_w$ matrix, which is a collection of estimated generation time series for individual PVs. The first term in the objective function (\ref{eq:overall_a}) reflects the difference between the estimated aggregate PV generation, $\hat{\pmb{G}}_w$, and the weighted summation of individual PV's estimated generations, $\mathbf{G}_e * \mathbf{K} * \pmb{1}$. The second term in the objective function (\ref{eq:overall_a}) considers the estimation errors of peak generations. $\lambda$ is a tuning parameter. $\pmb{\gamma}$ is an $N_w$-by-1 vector with non-negative elements, which reflect the errors of approximating $G_{w,m,i}$ as $D_{w,i}$, as shown in (\ref{eq:G_{w,m,i}}). The second $\pmb{1}$ is a $T$-by-1 vector of ones. $\hat{\pmb{G}}_{w,m} = [\hat{G}_{w,m,1},...,\hat{G}_{w,m,N_w}]^\mathsf{T}$ denotes an $N_w$-by-1 vector of the estimated generation peaks for all PVs. $(\hat{\pmb{G}}_{w,m}+\pmb{\gamma})$ denotes the corrected generation peaks with consideration of estimation errors. $\pmb{1}* (\hat{\pmb{G}}_{w,m} + \pmb{\gamma})^\mathsf{T}$ produces a $T$-by-$N_w$ matrix , in which each column contains the same element. Constraint (\ref{eq:overall_b}) ensures that the estimated generation time series for each PV is smaller than its estimated peak generation. $\pmb{0}$ is an $N_w$-by-1 vector of zeros. $P_0$ denotes the maximum error of approximating $G_{w,m,i}$ as $D_{w,i}$ for individual PVs. The third $\pmb{1}$ is an $N_w$-by-1 vector of ones. Constraint (\ref{eq:overall_d}) ensures that the estimation errors for individual PVs are non-negative and smaller than an upper bound. The reason for constraining the elements of $\pmb{\gamma}$ as non-negative is that $D_{w,i}$ typically under-estimates $G_{w,m,i}$, as shown in Fig. \ref{fig:noc_min_P_vs_diu_min_P}. 

The optimization process represented in (\ref{eq:overall}) is a convex quadratic programming problem, thus, we can obtain a unique solution for $\mathbf{K}$, i.e., $\mathbf{K}^*=[\pmb{K}_1^*, ..., \pmb{K}_{N_w}^*]$. Then, the estimated generation time series for the $i$'th PV, $\hat{\pmb{G}}_{w,i}=\{\hat{G}_{w,i}(t)\}$, $t=1,...,T$, can be computed as:
\begin{equation}  \label{eq:hat_G_w_i}
\hat{\pmb{G}}_{w,i} = \mathbf{G}_e * \pmb{K}_i^*, \quad i=1,...,N_w.
\end{equation}

Then, the estimated native demand time series for the $i$'th customer, $\hat{\pmb{P}}_{w,i}=\{\hat{P}_{w,i}(t)\}$, $t=1,...,T$, can be computed as:
\begin{equation}  \label{eq:hat_P_w_i}
\hat{\pmb{P}}_{w,i} = \pmb{P}_{w,i}' + \hat{\pmb{G}}_{w,i}, \quad i=1,...,N_w.
\end{equation}
where, $\pmb{P}_{w,i}'=\{{P}_{w,i}'(t)\}$, $t=1,...,T,$ denotes the known net demand time series recorded by smart meter for the $i$'th customer with PVs.

\begin{figure}
\centering
\includegraphics[width=1\linewidth]{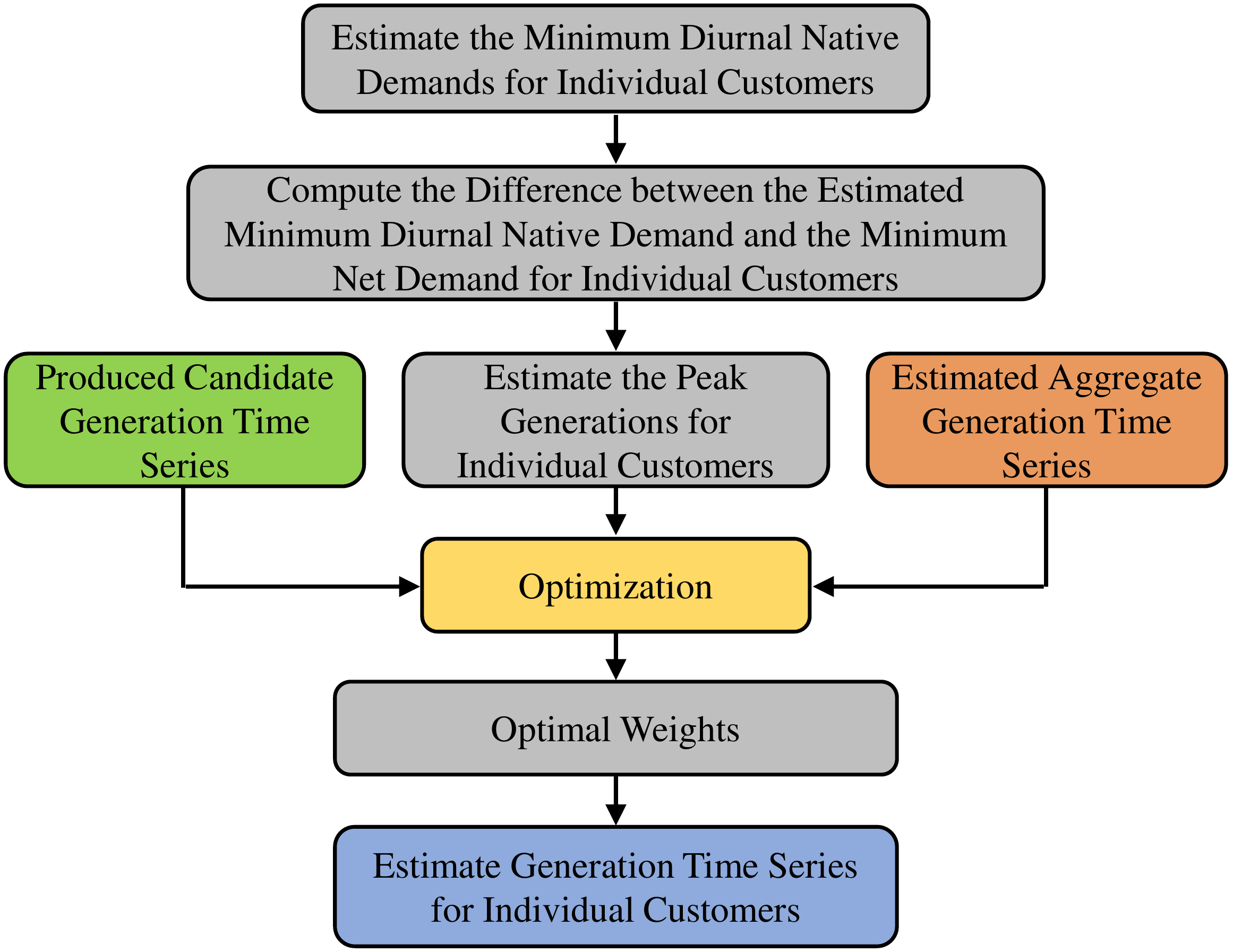}
\caption{\hl{Detailed steps of the individual customer-level BTM PV generation estimation.}}
\label{fig:customer_layer}
\end{figure}

Note that (\ref{eq:overall}) can be solved for a selected window. The window size, $T$,  can impact estimation accuracy and runtime, which will be examined in the Case Study Section. The detailed steps for estimating customer-level PV generation are illustrated in Fig. \ref{fig:customer_layer}. 

%%**************************%%
%                            %
%          Case Study        %
%                            %
%% *************************%%
\section{Case Study}\label{sec:case_study}
In this section, the proposed two-layer BTM solar power and native demand estimation approach is verified using real PV generation and native demand data.

\subsection{Dataset Description}
The hourly native demand and PV generation data used in this paper are from a public dataset \cite{data_source}. The time range of native demand and solar power is one year. This dataset contains a total number of 100 customers with PVs and 115 customers without PVs. For the customers with PVs, the net demand is obtained by subtracting PV generation from native demand.

\subsection{Aggregate-level BTM PV Generation Estimation Validation}
\begin{figure}[htbp]
\centering
\subfloat[Aggregate PV generation\label{sfig:real_esti_aggre_G_curves}]{
\includegraphics[width=0.75\linewidth]{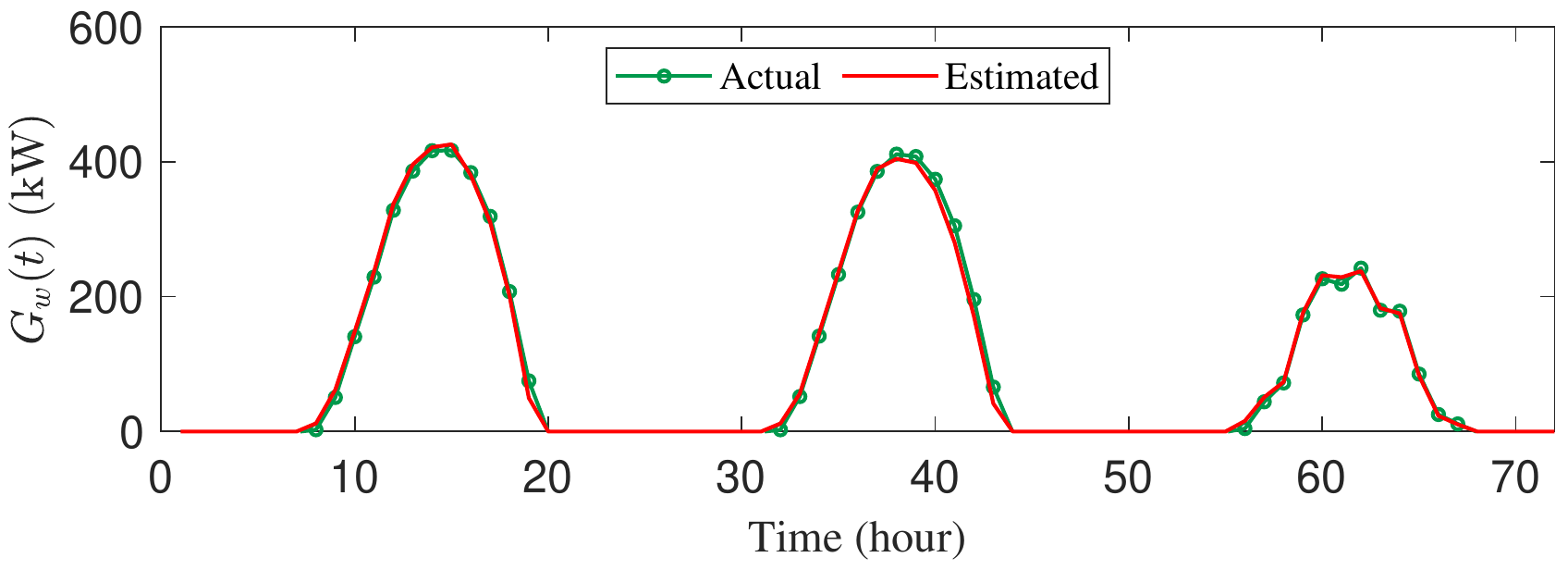}
}
\hfill
\subfloat[Aggregate native demand\label{sfig:real_esti_aggre_P_curves}]{
\includegraphics[width=0.75\linewidth]{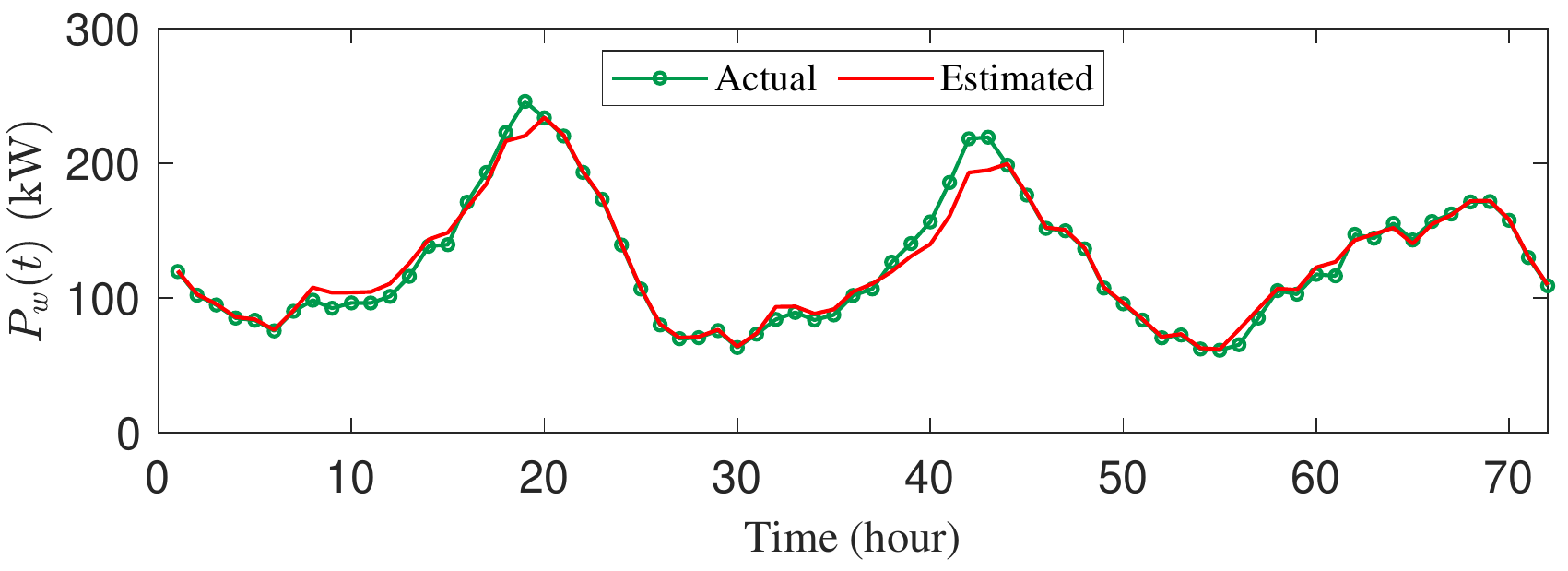}
}
\caption{\hl{Three-day actual and estimated aggregate PV generation and native demand curves.}}
\label{fig:aggre_real_esti_curves}
\end{figure}

Fig. \ref{fig:aggre_real_esti_curves} shows three-day actual and estimated aggregate PV generation/native demand curves. It can be seen that the estimated curves can accurately follow the actual curves. \hl{To quantitatively evaluate the estimation accuracy, we compute the mean absolute percentage error (MAPE) as follows:}
% \begin{subequations}  \label{eq:MAPE_agre}
% \begin{equation}   \label{eq:MAPE_agre_G}
% MAPE_G = \frac{100\%}{N_d}\cdot \sum_{t \in I_d}^{} \Bigg| \frac{\hat{G}_w(t)-{G}_w(t)}{G_{w,m}}  \Bigg|, 
% \end{equation}    
% \begin{equation}    \label{eq:MAPE_agre_P}
% MAPE_P = \frac{100\%}{N_d}\cdot \sum_{t \in I_d}^{} \Bigg| \frac{\hat{P}_w(t)-{P}_w(t)}{P_{w,m}}  \Bigg|, 
% \end{equation}
% \end{subequations}

\begin{equation}  \label{eq:MAPE_agre}
MAPE = \frac{100\%}{N_d} \sum_{t \in I_d}^{} \Bigg| \frac{\hat{Y}_{w}(t)-{Y}_{w}(t)}{Y_{w,m}}  \Bigg|,
\end{equation}
\hl{where, $\hat{Y}_{w}(t)$ represents $\hat{G}_w(t)$ or $\hat{P}_w(t)$. ${Y}_{w}(t)$ represents ${G}_w(t)$ or ${P}_w(t)$. $Y_{w,m}$ represents $G_{w,m}$ or $P_{w,m}$, where $G_{w,m}$ and $P_{w,m}$ denote the actual peaks of PV generation and native demand, respectively.} $I_d$ denotes the set of daytime hours. $N_d$ denotes the total number of hours in $I_d$.

\hl{To comprehensively evaluate the performance of our approach, we also compute the mean squared error (MSE) and coefficient of variation (CV):}
\begin{equation}  \label{eq:MSE_agre}
MSE = \frac{1}{N_d} \sum_{t \in I_d}^{} \big(\hat{Y}_{w}(t)-{Y}_{w}(t) \big)^2,
\end{equation}
\begin{equation}  \label{eq:CV_agre}
CV = \frac{\sigma}{\mu},
\end{equation}
\hl{where, }
\begin{subequations}  \label{eq:mu_sigma_agre}
\begin{equation}   \label{eq:mu_agre}
\mu=\frac{1}{N_d} \sum_{t \in I_d}^{} (\hat{Y}_{w}(t)-{Y}_{w}(t) ), 
\end{equation}    
\begin{equation}    \label{eq:sigma_agre}
\sigma = \sqrt{\frac{1}{N_d-1}\sum_{t \in I_d}^{} \big((\hat{Y}_{w}(t)-{Y}_{w}(t)) -\mu \big)^2}.
\end{equation}
\end{subequations}

The computed $MAPE$'s for PV generation and native demand are 1.21\% and 1.28\%, respectively. \hl{The computed $MSE$'s for PV generation and native demand are about 58.09. Note that the actual peaks for the PV generation and native demand are 462.5 and 437.1 kW, respectively. The computed $CV$'s for PV generation and native demand are about -3.48. The above error metrics reflect the high accuracy of our proposed approach. } 

\subsection{Customer-level BTM PV Generation Estimation Validation}
\subsubsection{Estimation Performance}
\begin{figure}[htbp]
\centering
\subfloat[PV generation\label{sfig:real_esti_indi_G_curves}]{
\includegraphics[width=0.75\linewidth]{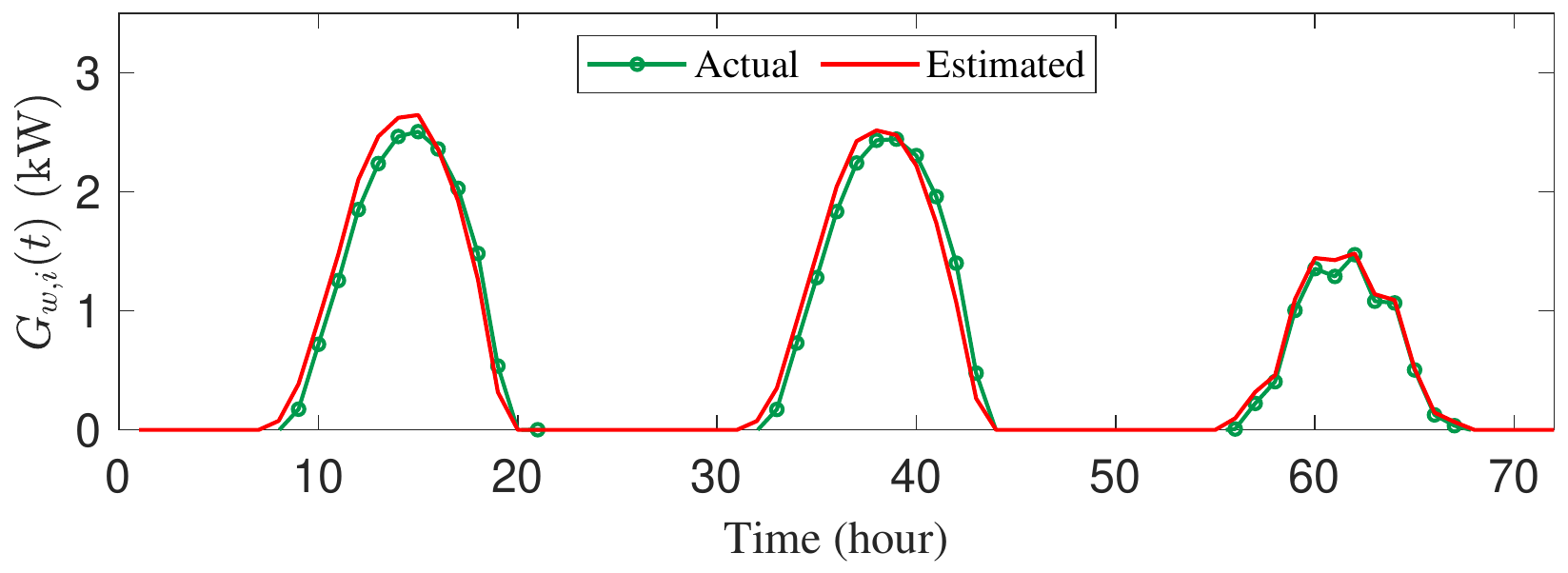}
}
\hfill
\subfloat[Native demand\label{sfig:real_esti_indi_P_curves}]{
\includegraphics[width=0.75\linewidth]{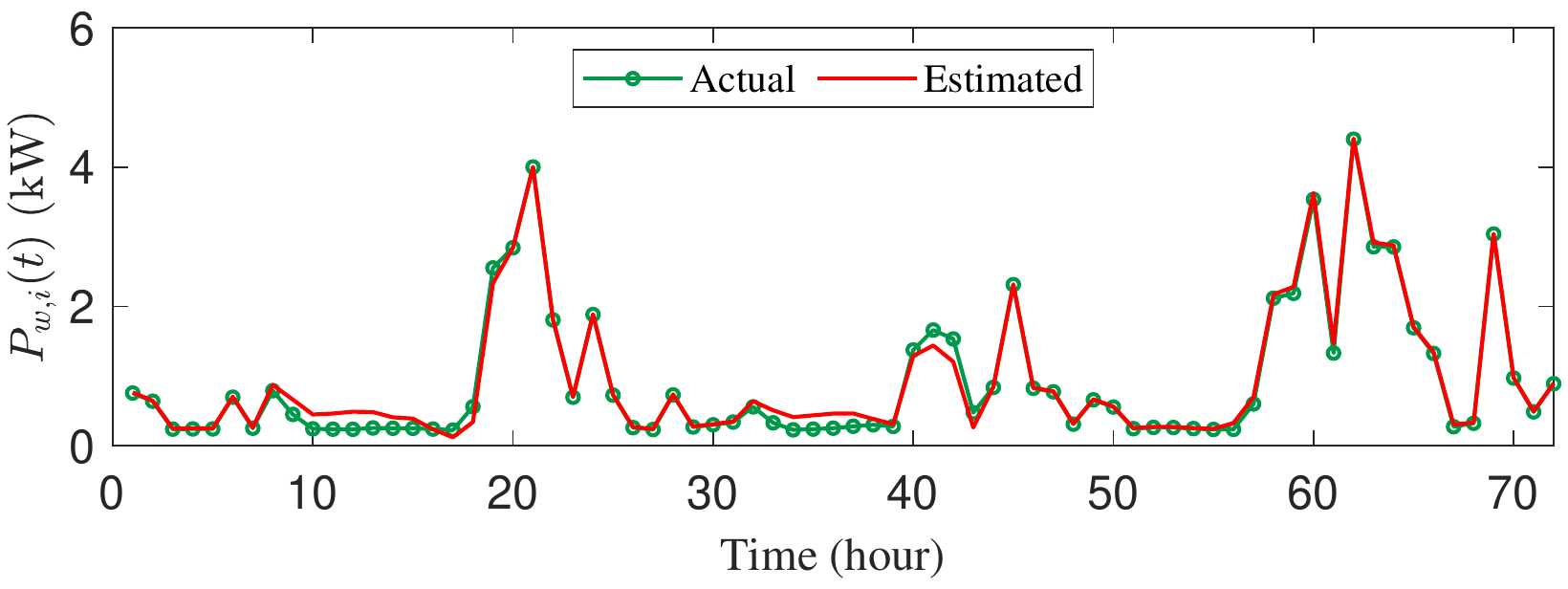}
}
\caption{\hl{Three-day actual and estimated PV generation and native demand curves for an example customer with PV.}}
\label{fig:indi_real_esti_curves}
\end{figure}

Fig. \ref{fig:indi_real_esti_curves} shows three-day actual and estimated PV generation and native demand curves for an example customer with PV. We can see that the estimated curves can accurately fit the actual curves. To comprehensively examine the performance of our approach, we compute the $MAPE$ for all customers with PVs. Specifically, the $MAPE$'s for the $i$'th customer are computed \hl{as follows:}
% \begin{subequations}  \label{eq:MAPE_indi}
% \begin{equation}   \label{eq:MAPE_indi_G}
% MAPE_{G,i} = \frac{100\%}{N_d}\cdot \sum_{t \in I_d}^{} \Bigg| \frac{\hat{G}_{w,i}(t)-{G}_{w,i}(t)}{G_{w,m,i}}  \Bigg|, 
% \end{equation}    
% \begin{equation}    \label{eq:MAPE_indi_P}
% MAPE_{P,i} = \frac{100\%}{N_d}\cdot \sum_{t \in I_d}^{} \Bigg| \frac{\hat{P}_{w,i}(t)-{P}_{w,i}(t)}{P_{w,m,i}}  \Bigg|, 
% \end{equation}
% \end{subequations}

\begin{equation}  \label{eq:MAPE_indi}
MAPE_{i} = \frac{100\%}{N_d} \sum_{t \in I_d}^{} \Bigg| \frac{\hat{Y}_{w,i}(t)-{Y}_{w,i}(t)}{Y_{w,m,i}}  \Bigg|
\end{equation}
\hl{where $Y_{w,i}(t)$ represent $G_{w,i}(t)$ or $P_{w,i}(t)$, $\hat{Y}_{w,i}(t)$ represent $\hat{G}_{w,i}(t)$ or $\hat{P}_{w,i}(t)$, and $Y_{w,m,i}$ represent $G_{w,m,i}$ or $P_{w,m,i}$. $G_{w,m,i}$ and $P_{w,m,i}$ denote the actual generation and native demand peaks for the $i$'th customer, respectively. We also compute the $MSE$ and $CV$ for each PV-installed customer:}
\begin{equation}  \label{eq:MSE_indiv}
MSE_i = \frac{1}{N_d} \sum_{t \in I_d}^{} \big(\hat{Y}_{w,i}(t)-{Y}_{w,i}(t) \big)^2,
\end{equation}
\begin{equation}  \label{eq:CV_indiv}
CV_i = \frac{\sigma_i}{\mu_i},
\end{equation}
\hl{where, }
\begin{subequations}  \label{eq:mu_sigma_indiv}
\begin{equation}   \label{eq:mu_indiv}
\mu_i=\frac{1}{N_d} \sum_{t \in I_d}^{} (\hat{Y}_{w,i}(t)-{Y}_{w,i}(t) ), 
\end{equation}    
\begin{equation}    \label{eq:sigm_indiv}
\sigma_i = \sqrt{\frac{1}{N_d}\sum_{t \in I_d}^{} \big((\hat{Y}_{w,i}(t)-{Y}_{w,i}(t)) -\mu_i \big)^2}.
\end{equation}
\end{subequations}

Table \ref{tbl:MAPE_MSE_CV} \hl{summarises the empirical cumulative distribution functions (CDFs) for the estimation $MAPE$, $MSE$, and $CV$, which are constructed using all the computed $MAPE$'s, $MSE$'s, and $CV$'s, respectively.} As can be seen, for the estimated hourly PV generation, 70\% of the $MAPE$'s are less than 6.38\%. Regarding the estimated hourly native demand, 70\% of the $MAPE$'s are less than 3.67\%. This effectively verifies the estimation accuracy of our proposed approach. \hl{We also provide the percentiles of $MSE$ and $CV$ based on all the PV-installed customers' generation and native demand estimates, which can more comprehensively evaluate the performance of our approach. }

\begin{table}[htbp]
\centering
	\renewcommand{\arraystretch}{1.5}
	\setlength{\tabcolsep}{4.8pt}
	\caption{\hl{Empirical CDF of Estimation Error Metrics}}\label{tbl:MAPE_MSE_CV}
	\begin{tabular}{cccccc}
	    \toprule[1pt]
		Empirical CDF & 0.1 & 0.2 & 0.5 & 0.7 & 0.9  \\
	    \hline
        $MAPE$ of $\hat{G}$ (\%) & 2.84  & 4.05 & 4.96  & 6.38  & 8.80 \\
        $MAPE$ of $\hat{P}$ (\%) & 1.63  & 2.15 & 2.80  & 3.67  & 4.92\\
        $MSE$ of $\hat{G}$ & 0.04  & 0.06 & 0.10  & 0.19  & 0.33 \\
        $MSE$ of $\hat{P}$ & 0.03  & 0.05 & 0.09  & 0.18  & 0.29 \\
        $CV$ of $\hat{G}$ & -11.80  & -5.13 & -2.60  & 2.37  & 16.12 \\
        $CV$ of $\hat{P}$ & -11.30  & -4.65 & -2.59  & 1.77  & 10.90 \\
        \bottomrule[1pt]
	\end{tabular}
\end{table}

Note that the above results are obtained under the conditions that (1) five produced candidate generation curves are employed ($N_e=5$), (2) the tuning parameter in (\ref{eq:overall_a}) is 100 ($\lambda=100$), and (3) the optimization process specified in (\ref{eq:overall}) is executed for individual windows with a time length of one month ($T=720$ hours, the entire year is divided into 12 windows).

\subsubsection{Testing the Candidate Generation Curves}
\begin{figure}
\centering
\includegraphics[width=0.75\linewidth]{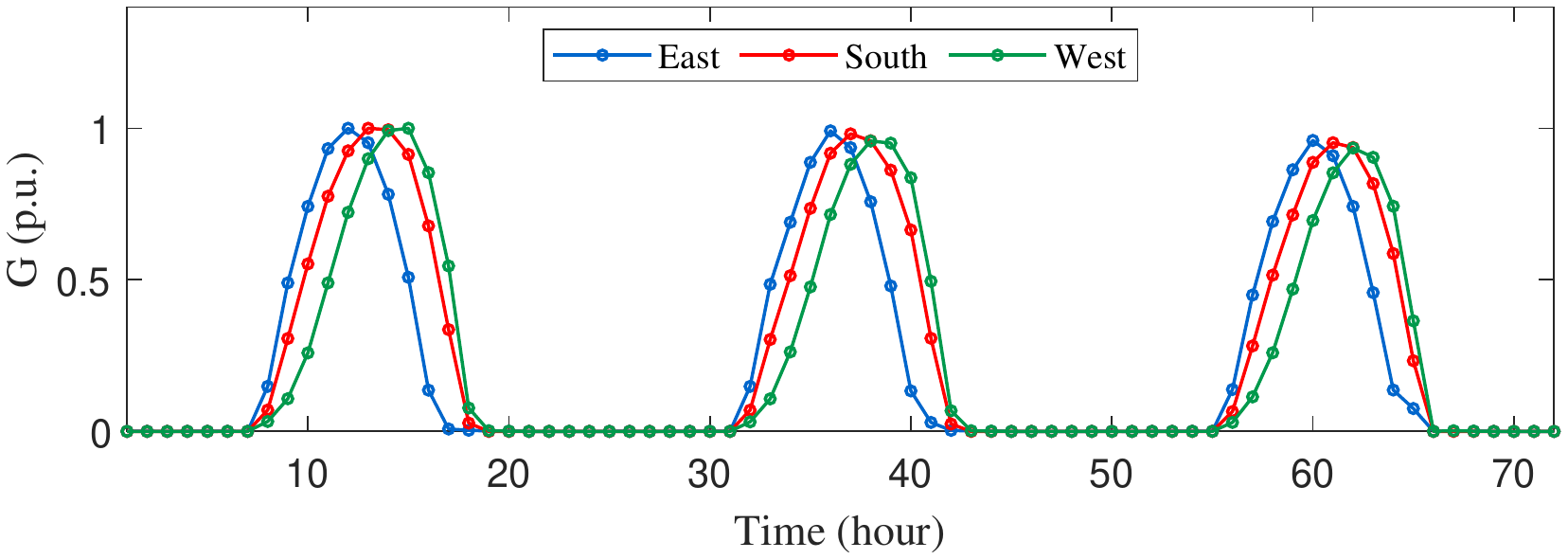}
\caption{Three-day produced candidate generation curves corresponding to three typical azimuths, i.e., east, south, and west.}
\label{fig:norm_candid_generation_curves}
\end{figure}

As elaborated in Section \ref{sec:indiv_level_shape}, diverse candidate generation curves are produced for representing the unknown BTM generation. Thus, it is of interest to examine the effectiveness of producing candidate curves. Fig. \ref{fig:norm_candid_generation_curves} shows three produced candidate generation curves corresponding to three typical azimuths, i.e., east, south, and west, respectively. We can observe that compared to the generation curve corresponding to the south, the produced curve corresponding to the east is ``left-skewed'', and the produced curve corresponding to the west is ``right-skewed''. Therefore, the produced curves demonstrate diversity, which is consistent with our observation on real PV generation curves shown in Fig. \ref{fig:nrm_agg_indi_gener_curves_1}. 

In addition, we have also quantitatively examined the effectiveness of producing diverse candidate generation curves. Specifically, we test the impact of the number of candidate generation curves, i.e., we solve (\ref{eq:overall}) separately for three cases with different numbers of candidate curves: (I) one candidate generation curve corresponding to the azimuth of south; (II) three candidate generation curves corresponding to the east, south, and west, respectively; and (III) five candidate generation curves corresponding to the east, southeast, south, southwest, and west, respectively. The other conditions for the three cases are the same: $\lambda=100$ and $T=720$ hours. To evaluate the impact of candidate number, we compute the average $MAPE$ over all PVs' $MAPE$'s obtained from (\ref{eq:MAPE_indi}). The results are summarized in Table \ref{tbl:MAPE_impact_of_candidate}. We can see that as the candidate number increases, the estimation error decreases, and the execution time increases. In addition, the $MAPE$ for Case I is relatively greater than Case II and III, and Case II and Case III provide nearly identical $MAPE$'s. This is because three candidate curves - corresponding to the east, south, and west - can comprehensively represent the unknown BTM generation curve; adding extra candidate curves simply result in a slight accuracy improvement. 

\begin{table}[htbp]
\centering
	\renewcommand{\arraystretch}{1.5}
	\setlength{\tabcolsep}{4.8pt}
	\caption{Impact of Candidate Generation Curves}\label{tbl:MAPE_impact_of_candidate}
	\begin{tabular}{cccc}
	    \toprule[1pt]
		 Case  & I & II & III   \\
	    \hline
        Average $MAPE$ of $\hat{G}$ (\%) & 5.677  & 5.474 & 5.473  \\
        Average $MAPE$ of $\hat{P}$ (\%) & 3.924  & 3.086 & 3.086  \\
        Runtime (s)                      & 40     & 125   & 194  \\
        \bottomrule[1pt]
	\end{tabular}
\end{table}

\subsubsection{Testing the Tuning Parameter $\lambda$}
As discussed in Section \ref{sec:optimization_formulation}, $\lambda$ in (\ref{eq:overall}) reflects the confidence of estimating peak generations for individual PVs. One general principle for determining $\lambda$ is that the largest element in $\pmb{\gamma}$ is a couple of kilo-watts. In addition, the solutions for (\ref{eq:overall}) should not be sensitive to $\lambda$, i.e., (\ref{eq:overall}) should be robust to $\lambda$. To verify the robustness of our proposed approach, we solve (\ref{eq:overall}) based on different values of $\lambda$, and then compute the corresponding average $MAPE$'s for the estimated PV generation and native demand. Other conditions are that $T=720$ hours and five candidate generation curves - corresponding to the south, southeast, south, southwest, and west - are employed. The results show that for the $\lambda$'s ranging from 100 to 500 with an interval of 100, the average $MAPE$'s for PV generation and native demand do not change (5.47\% and 3.09\%). The invariant average $MAPE$'s demonstrate the robustness of our proposed approach.

% \begin{table}[htbp]
% \centering
% 	\renewcommand{\arraystretch}{1.5}
% 	\setlength{\tabcolsep}{4.8pt}
% 	\caption{Testing the Tuning Parameter $\lambda$}\label{tbl:MAPE_impact_of_lambda}
% 	\begin{tabular}{cccccc}
% 	    \toprule[1pt]
% 		 $\lambda$  & 100 & 200 & 300  & 400 & 500  \\
% 	    \hline
%         Average $MAPE$ of $\hat{G}$ (\%) & 5.47  & 5.47 & 5.47 & 5.47 & 5.47  \\
%         Average $MAPE$ of $\hat{P}$ (\%) & 3.09  & 3.09 & 3.09 & 3.09 & 3.09  \\
%         \bottomrule[1pt]
% 	\end{tabular}
% \end{table}

\subsubsection{Testing the Window Size $T$}
Since our proposed approach can be conducted for each divided window, it is of importance to examine the impact of window size on estimation accuracy. To do this, we perform our approach for windows with different lengths and then compute the estimation $MAPE$. In Table \ref{tbl:MAPE_impact_of_T}, it can be seen that the average $MAPE$ decreases as $T$ increases. This is because for a wider window, the probability for the minimum diurnal native demand, $\underbar{$P$}_{w,d,i}$, equaling the minimum nocturnal native demand, $\underbar{$P$}_{w,n,i}$, is larger. Thus, we have a smaller estimation error for $\underbar{$P$}_{w,d,i}$, as seen in (\ref{eq:diu_min_P_i}). Then, based on (\ref{eq:D_{w,i}}) and (\ref{eq:G_{w,m,i}}), it can be seen that the smaller estimation error for $\underbar{$P$}_{w,d,i}$ results in a more accurate $D_{w,i}$, which then brings a more accurate estimate for $G_{w,m,i}$. Finally, more accurate peak generation estimates result in smaller estimation errors for the PV generation and native demand time series.

\begin{table}[htbp]
\centering
	\renewcommand{\arraystretch}{1.5}
	\setlength{\tabcolsep}{4.8pt}
	\caption{Impact of Window Size $T$}\label{tbl:MAPE_impact_of_T}
	\begin{tabular}{ccccc}
	    \toprule[1pt]
		 $T$ (month)  & 1 & 2 & 3  & 4   \\
	    \hline
        Average $MAPE$ of $\hat{G}$ (\%) & 5.47  & 5.30 & 5.18 & 5.08  \\
        Average $MAPE$ of $\hat{P}$ (\%) & 3.09  &2.99 & 2.92 & 2.87  \\
        \bottomrule[1pt]
	\end{tabular}
\end{table}

\subsection{Performance Comparison}
This paper compares our proposed approach with previous works from two perspectives, qualitatively and quantitatively. 
\subsubsection{Qualitative Analysis}
From a qualitative point of view, one primary advantage of our approach is that it does not require meteorological data and solar generation exemplars. For the aggregate level, our approach can perform PV generation estimation by only using recorded net demand data. For the customer level, our approach can also work by only relying on recorded smart meter data, although leveraging PVWatts Calculator's generated data can improve the estimation accuracy. 
\subsubsection{Quantitative Comparison}
For the customer level, we have also compared our approach with previous works. Specifically, we focus on comparing our approach with the method presented in \cite{Fankun_Bu_1} and \cite{nan_peng_yu_2}, which demonstrate better performance compared to previous works. Table \ref{tbl:MAPE_comparison} summarizes the computed $MAPE$'s for our approach and the compared approach. Note that the average $MAPE$'s for our approach have lower and upper bounds because the considered window size, $T$, ranges from one month to four months. As can be seen, the approach in \cite{Fankun_Bu_1} demonstrates a similar estimation accuracy as our approach does. However, our approach does not require solar exemplars, which makes it more independent and practical.\hl{ The approach in} \cite{nan_peng_yu_2}\hl{ employs a statistical model and a physical model to represent the native load and the PV generation, respectively. Table }\ref{tbl:MAPE_comparison}\hl{ shows that our approach has a better performance than the approach in }\cite{nan_peng_yu_2}\hl{ in terms of the average $MAPE$.}

\begin{table}[htbp]
\centering
	\renewcommand{\arraystretch}{1.5}
	\setlength{\tabcolsep}{4.8pt}
	\caption{\hl{Average $MAPE$ (\%) Comparison}}\label{tbl:MAPE_comparison}
	\begin{tabular}{cccccc}
	    \toprule[1pt]
		 Approaches  & Our Approach & \hl{Approach in} \cite{nan_peng_yu_2}  & Approach in \cite{Fankun_Bu_1} \\
	    \hline
        $\hat{G}$  & [5.08, 5.47]  &  7.38 & 5.24    \\
        $\hat{P}$ & [2.87, 3.09]  & 9.94 & 2.95    \\
        \bottomrule[1pt]
	\end{tabular}
\end{table}

\subsection{\hl{Robustness against Measurement and Communication Noises}}
\hl{To test the robustness of our proposed approach, we add measurement and communication noises to the net demand measurements of customers with PVs and the native demand measurements of customers without PVs. For the measurement noise, we consider the Class 0.5 (having $\pm$0.5\% error) specified by ANSI C12.20. For the communication noise, we test five different packet loss rates considering that the packet loss rate depends on the communication bandwidth and data volume. For example, we purposely change 1\% of the measurements to zero to achieve a 1\% packet loss rate. To comprehensively evaluate our approach's performance, we set up five cases: Case 1 - 1\% measurement lost + 0.5\% random noise, Case 2 - 2\% measurement lost + 0.5\% random noise, Case 3 - 3\% measurement lost + 0.5\% random noise, Case 4 - 4\% measurement lost + 0.5\% random noise, and Case 5 - 5\% measurement lost + 0.5\% random noise. Then, we apply our approach to the above five cases and compute the $MAPE$ for evaluating the robustness. The results are summarized in Table} \ref{tbl:MAPE_impact_of_noise_aggre} and \ref{tbl:MAPE_impact_of_noise_indiv}. \hl{We can observe that the $MAPE$'s slowly increase while the noise level increases, demonstrating the robustness of our approach.}
\begin{table}[htbp]
\centering
	\renewcommand{\arraystretch}{1.5}
	\setlength{\tabcolsep}{4.8pt}
	\caption{\hl{Aggregate-level Estimation $MAPE$ (\%)}}\label{tbl:MAPE_impact_of_noise_aggre}
	\begin{tabular}{ccccccc}
	    \toprule[1pt]
		   & W/O noise & Case 1 & Case 2 & Case 3 & Case 4 & Case 5   \\
	    \hline
        $\hat{G}$ & 1.21  & 1.17 & 1.22 & 1.38 & 1.53 & 1.73 \\
        $\hat{P}$ & 1.28  & 1.28 & 1.33 & 1.43 & 1.58 & 1.76\\
        \bottomrule[1pt]
	\end{tabular}
\end{table}

\begin{table}[htbp]
\centering
	\renewcommand{\arraystretch}{1.5}
	\setlength{\tabcolsep}{4.8pt}
	\caption{\hl{Average Customer-level Estimation $MAPE$ (\%)}}\label{tbl:MAPE_impact_of_noise_indiv}
	\begin{tabular}{ccccccc}
	    \toprule[1pt]
		   & W/O noise & Case 1 & Case 2 & Case 3 & Case 4 & Case 5   \\
	    \hline
        $\hat{G}$ & 5.47  & 5.84 & 5.86 & 5.64 & 5.54 & 5.62 \\
        $\hat{P}$ & 3.09  & 3.53 & 3.68 & 3.62 & 3.63 & 3.80\\
        \bottomrule[1pt]
	\end{tabular}
\end{table}

\subsection{\hl{Limitations of the Proposed Approach}}
\hl{Every method has its limitations, and there is no omnipotent method that can apply to all cases. The limitation of our proposed approach is that it requires time-series smart meter data with a temporal granularity that can distinguish daytime and nighttime. This is because our approach innovatively utilizes the temporal correlation between the aggregate \textit{nocturnal} native demand and the aggregate \textit{diurnal} native demand. Under this condition, only having access to the monthly demands of those PV-installed customers brings challenges to our approach because it cannot split the monthly demand into two parts, the diurnal and nocturnal demands, for computing the nocturnal native demand ratio. We intend to address this challenge in our future work.}

%%**************************%%
%                            %
%          Conclusion        %
%                            %
%% *************************%%
\section{Conclusion}\label{sec:conclusion}
This paper is dedicated to proposing an independent and practical BTM solar power/native demand estimation approach. Our proposed approach contains two interconnected layers. The aggregate level leverages the spatial correlation of native demand to perform the aggregate PV generation/native demand estimation. The customer level utilizes the spatial correlation of PV generation to allocate the estimated aggregate PV generation/native demand to individual customers. The Case Study verifies that our approach can accurately estimate BTM PV generation/native demand, significantly enhancing distribution system observability and situation awareness. The numerical experiments also demonstrate that our approach does not require meteorological data and measured solar power exemplars. Therefore, our approach is more independent and thus is practical for utilities to implement.

\ifCLASSOPTIONcaptionsoff
  \newpage
\fi



%\bibliographystyle{IEEEtran}


% \end{thebibliography}

\bibliographystyle{IEEEtran}
% \bibliography{IEEEabrv,./bibtex/bib/ref}
\bibliography{IEEEabrv,ref}

\end{document}